\documentclass[onecolumn,preprintnumbers,amsmath,amssymb,9pt]{revtex4}
\usepackage{amsmath}
\usepackage{rotating}
\usepackage[ruled]{algorithm2e}
\usepackage{extarrows}
\usepackage{graphicx}% Include figure files
\usepackage{dcolumn}% Align table columns on decimal point
\usepackage{bm}% bold math
\usepackage[all]{xy}
\usepackage{indentfirst}
\usepackage{amsmath}
\usepackage{bbm}
\usepackage{multirow}
\usepackage{mathrsfs}
\usepackage{euscript}
\usepackage{amssymb}

\begin{document}

%%%%%%%%%%%%%%%%%%%%%%%%%%%%%%%%%%%%%%%%%%%%%%%%%%%%%%%%%%%%%%%%%%%%%%%%%%%%%
%\preprint{APS}%/123-QED}
%%%%%%%%%%%%%%%%%%%%%%%%%%%%%%%%%%%%%%%%%%%%%%%%%%%%%%%%%%%%%%%%%%%%%%%%%%%%%
\title{Fully Device-Independent Model on Quantum Networks}
%%%%%%%%%%%%%%%%%%%%%%%%%%%%%%%%%%%%%%%%%%%%%%%%%%%%%%%%%%%%%%%%%%%%%%%%%%%%%
 \author{Ming-Xing Luo
\\
Information Security and National Computing Grid Laboratory, Southwest Jiaotong University, Chengdu 610031, China
}

\begin{abstract}
Bell inequality can provide a useful witness for device-independent applications with quantum (or post-quantum) eavesdroppers. This feature holds only for single entangled systems. Our goal is to explore device-independent model for quantum networks. We firstly propose a Bell inequality to verify the genuinely multipartite nonlocality of connected quantum networks including cyclic networks and universal quantum computational resources for measurement-based computation model. This is further used to construct new monogamy relation in a fully device-independent model with multisource quantum resources. It is finally applied for multiparty quantum key distribution, blind quantum computation, and quantum secret sharing. The present model can inspire various large-scale applications on quantum networks in a device-independent manner.
\end{abstract}
%%%%%%%%%%%%%%%%%%%%%%%%%%%%%%%%%%%%%%%%%%%%%%%%%%%%%%%%%%%%%%%%%%%%%%%%%%%%%
%%%%%%%%%%%%%%%%%%%%%%%%%%%%%%%%%%%%%%%%%%%%%%%%%%%%%%%%%%%%%%%%%%%%%%%%%%%%%
%%%%%%%%%%%%%%%%%%%%%%%%%%%%%%%%%%%%%%%%%%%%%%%%%%%%%%%%%%%%%%%%%%%%%%%%%%%%%
\maketitle
%\linenumbers

The Bell theorem states that the statistics generated by local measurements on a two-spin entanglement cannot be generated by any classical local model under the locality and casualty assumptions. This provides an experimental method for verifying the so-called nonlocality of entanglement \cite{1,2}. To date, Bell theory has inspired many interesting applications in various areas, such as quantum information processing \cite{3}, quantum key distribution \cite{4,5,6,7,8} and randomness amplification \cite{9,10,11}.

Single entangled systems have experimental constraints in large-scale applications because of the limited coherence time and transmission distance. This inspires distributed settings in terms of quantum networks using various independent entangled systems \cite{12,13}. Compared with single entangled systems, it is great difficulty in characterizing multipartite correlations of quantum networks because of the independent assumption of sources and local joint measurement allowed for each party. Moreover, the participation of multiple parties in a general network may provide new opportunities for attackers. Thus, a natural problem is how to ensure secure information processing on general quantum networks? The main difficulty is that these multipartite quantum correlations can form non-convex semialgebraic sets \cite{14}. Several nonlinear Bell inequalities have recently been proposed for verifying the non-multilocality of special quantum networks with the assumption of source independence \cite{15,16,16a,16b,17,luo,18,18b,18c,18d,18e,18f}. However, so far there is no way to verify the multipartite nonlocality of general quantum networks, or no standard definition of genuinely multipartite nonlocality for quantum networks.

In most of secure tasks, such as quantum key distribution (QKD) \cite{5}, the trustworthiness of quantum devices according to certain specifications should be avoided in order to enable adversary (noise)-tolerant realizations \cite{6}. These so-called device-independent scenarios make only use of the statistics of measurement outcomes \cite{19,20,21}. Interestingly, the leaked information in the case of quantum  eavesdroppers (or the key rate of QKD) may be characterized by the violation of specific inequality \cite{6,7,8}, that is, the higher the violation is, the lower the information leakage for the outcomes of legal parties. Note that quantum devices may be correlated  by an untrusted producer (adversary or eavesdropper) in a device-independent model \cite{6}. This implies that the recent models \cite{22,23} are not device independent because of the independent assumption of sources. The independence of sources in these secure scenarios cannot be guaranteed by experiment improvements. Another reason is that an untrusted party in network scenarios may correlate the shared sources locally in secure applications. Hence, the question of how to construct a device-independent model remains an open problem for general quantum networks.

Our goal in this work is to propose a device-independent model for secure information processing in general quantum networks against quantum (or post-quantum) eavesdroppers. We first propose a new Bell inequality for verifying the genuinely multipartite nonlocality of quantum networks in the biseparable model \cite{26}. Compared with a recent result for ring-shaped networks \cite{24}, the proposed inequality provides the first Bell test for verifying general cyclic networks. This further implies a new feature  for characterizing the leaked information in device-independent tasks on quantum networks going beyond single entangled systems \cite{6,7,8,19,20,21} or device-independent models \cite{22,23}. It is then used to guarantee the security of various tasks, such as multipartite quantum key distribution, blind quantum computation with multiple servers, and quantum secret sharing using quantum networks. These results can inspire interesting applications on large-scale quantum networks in a device-independent manner.

\begin{figure}
\begin{center}
\resizebox{240pt}{120pt}{\includegraphics{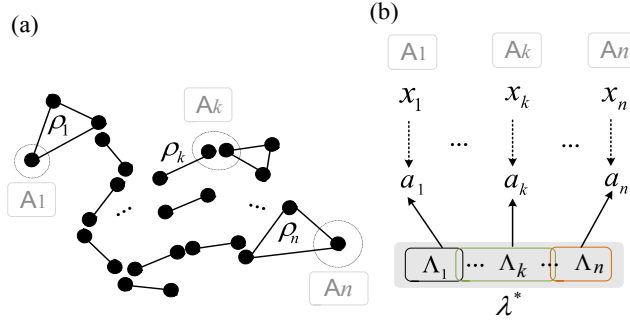}}
\end{center}
\caption{Quantum networks.(a) An $n$-partite quantum network ${\cal N}_q$ consisting of $m$ entangled states $\otimes_{i=1}^m\rho_i$. (b) A directed acyclic graph (DAG) \cite{25} of ${\cal N}_q$ consisting of $m$ independent random variables $\lambda_1, \cdots, \lambda_m$. In device-independent applications, a quantum (or post-quantum) eavesdropper can correlate shared variables $\lambda_{i}$s into new variable $\lambda^*$.}
\label{fig1}
\end{figure}

{\bf Genuinely multipartite nonlocality of quantum networks}. Consider an $n$-partite quantum network ${\cal N}_q$, as shown in Fig.\ref{fig1}(a), consisting of $m$ entangled states $\otimes_{i=1}^m\rho_i$. ${\cal N}_q$ is schematically represented by a directed acyclic graph (DAG) (Fig.\ref{fig1}(b)) \cite{25} with the same configuration as ${\cal N}_q$ using random variables $\lambda_1, \cdots, \lambda_m$. Different from the standard hidden variable model \cite{1,2}, a quantum (or post-quantum) eavesdropper is able to correlate all distributed variables $\lambda_i$s into a new variable $\lambda^*$ (Fig.\ref{fig1}(b)). Unfortunately, this local model is insufficient for characterizing device-independent applications including quantum secret sharing, where some sharers may be eavesdropper. In this case, we consider the genuinely multipartite nonlocality of quantum networks in the biseparable model \cite{26}. Let $I=\{s_1,\cdots, s_i\}\subset \{1, \cdots, n\}$ and $\overline{I}=\{j|j\not\in I, 1\leq j\leq n\}$ be a bipartition of $\{1, \cdots, n\}$. An $n$-partite state $\rho$ on Hilbert space $\otimes_{i=1}^n{\cal H}_{i}$ is biseparable \cite{26} if it has the following decomposition
\begin{eqnarray}
\rho=\sum_{I}p_{I}\varrho_{I,\overline{I}}
\end{eqnarray}
where $\{p_{I}\}$ is a probability distribution, $\varrho_{I,\overline{I}}$ denotes separable state on Hilbert space ${\cal H}_{I}\otimes {\cal H}_{\overline{I}}$, ${\cal H}_{I}=\otimes_{j\in I}{\cal H}_j$ and ${\cal H}_{\overline{I}}=\otimes_{j\in \overline{I}}{\cal H}_j$.

Let $M_{x_k(I)}$ denote dichotomic observable of the observer $\textsf{A}_k$ for input index $x_k(I)$, where $x_k$ depends on the set $I$. Denote $\textbf{M}_{I}=\prod_{k\in I}M_{x_k(I)}$ and  $\textbf{M}_{\overline{I}}=\prod_{k\in{}\overline{I}}M_{x_{k}(\overline{I})}$. Our first result is to prove that any biseparable state satisfies the following Bell inequality \cite{SI}:
\begin{eqnarray}
\sum_{i=1}^{\lfloor\frac{n}{2}\rfloor}{\cal L}_i\leq 2^{n}\sqrt{2}-4\sqrt{2}+2
\label{eq5}
\end{eqnarray}
where ${\cal L}_i$ denotes the summation of multipartite CHSH-type quantities, that is, ${\cal L}_i=\sum_{|I|=i}CH_{I}$ with $CH_{I}=(\textbf{M}_{I}+
\hat{\textbf{M}}_{I})\textbf{M}_{\overline{I}}
+(\textbf{M}_{I}-\hat{\textbf{M}}_{I})
\hat{\textbf{M}}_{\overline{I}}$, $\textbf{M}_{S}$ and $\hat{\textbf{M}}_{S}$ are observables depending on the set $S=I$ or $\overline{I}$, and $\lfloor{}x\rfloor$ denotes the maximum integer no more than $x$. The largest bound for quantum networks is $2^{n}\sqrt{2}-2\sqrt{2}$ \cite{SI}.

The inequality (\ref{eq5}) can be regarded as Svetlichny-type inequality \cite{26} with at most eight inputs \cite{SI}. The present inequality (\ref{eq5}) may reduce to Svetlichny inequality with specific settings \cite{26}. It is useful for verifying the genuinely multipartite nonlocality of connected quantum networks consisting of EPR states \cite{2} and GHZ states \cite{30} or noisy quantum networks (including two examples of triangle networks) \cite{Wrner,SI}. The present nonlocality is stronger than the non-multilocality \cite{15,16,16a,16b,17,luo,18,18b,18c,18d,18e,18f} or the nonlocality \cite{24,30,31,32}.

\begin{figure}
\begin{center}
\resizebox{240pt}{160pt}{\includegraphics{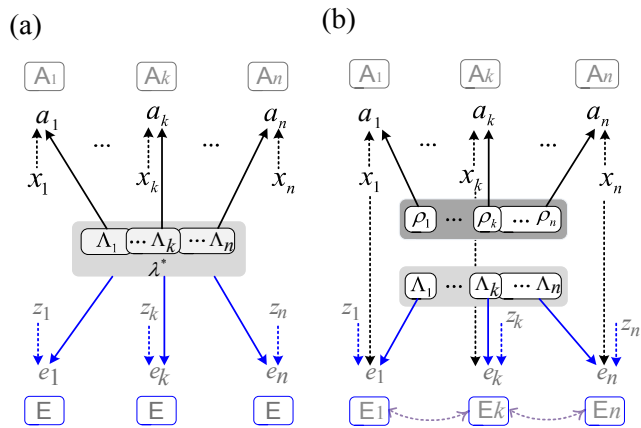}}
\end{center}
\caption{(a) A device-independent model on an $n$-partite network ${\cal N}$ consisting of $m$ sources $\lambda_1, \cdots, \lambda_m$. A quantum (post-quantum) eavesdropper $\textsf{E}$ holds some systems correlated with the shared sources. $z_i$ and $e_i$ denote the respective input and outcome of eavesdropper for recovering the output $a_i$. For a device-independent model, $\textsf{E}$ may correlate $\lambda_i$s into new source $\lambda^*$. (b)  Consider a quantum network ${\cal N}_q$, where $\textsf{A}_i$ shares a system $\rho_i$ consisting of some EPR states and GHZ states with $\textsf{A}_j$, $i=1, \cdots, k$, and $j=k+1,\cdots, n$. $\textsf{E}_i$s are eavesdroppers sharing random variables $\Lambda_i$ consisting of $\lambda_j$ with the same network configuration as ${\cal N}_q$. $e_i$ is the output of $\textsf{E}_i$ conditional on the inputs $z_i$ and $x_i$ and the variables $\Lambda_i$. $\textsf{E}_i$s want to classically simulate quantum correlations derived from local measurements on ${\cal N}_q$.}
\label{fig2}
\end{figure}

\textbf{A device-independent model on quantum networks}. For single entangled states, the violation of a certain Bell inequality allows for device-independent information processing against quantum (post-quantum) eavesdroppers \cite{5,6,7,8,9,10,11,19,20,21,26}. Our goal here is to propose a device-independent model using quantum networks based on the inequality (\ref{eq5}). Consider a connected network ${\cal N}$ (Fig.\ref{fig2}(a)). For a device-independent model, eavesdropper may hold local systems correlated with some sources in order to recover private information such as outcomes.

Denote the variation distance of two probability distributions $\{p(x)\}$ and $\{q(x)\}$ as: $D(p,q)=\frac{1}{2}\sum_x|p(x)-q(x)|$. The predictive power of an eavesdropper to learn the outcomes of legitimate parties satisfies \cite{SI}:
\begin{eqnarray}
D_e\leq 2n-\frac{\varpi_q-2\sqrt{2}
(\alpha-\lfloor{}\alpha\rfloor)}{2\lfloor{}\alpha\rfloor}
\label{eq6}
\end{eqnarray}
where $D_e=D(\prod_{i=1}^nP(e_i|a_i;\textbf{x},z_i), \prod_{i=1}^nP(e_i|z_i))$, $\alpha=\frac{1}{n}(2^n-1)$, $\textbf{x}=x_1\cdots{}x_n$,  $\varpi_q=\sum_{i=1}^{\lfloor\frac{n}{2}\rfloor}{\cal L}_i$, and ${\cal L}_i$ is defined in the inequality (\ref{eq5}) with the estimated quantum correlations. The inequality (\ref{eq6}) provides a monogamy relation for the information leakage in a device-independent model on quantum networks.

If all inputs are achievable for eavesdroppers as shown in Fig.\ref{fig2}(b), the quantum correlations derived from $\mathcal{N}_q$ may be classically simulated with shared randomness and finite classical communication \cite{SI}. This partially answers a recent conjecture \cite{36}. It inspires special restraints on the measurements to enhance the security for cryptographic applications.

\begin{figure}
\begin{center}
\resizebox{240pt}{120pt}{\includegraphics{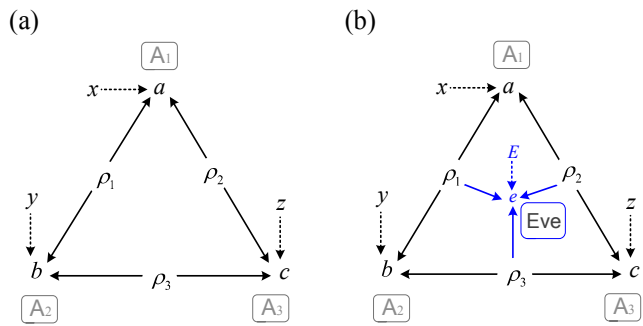}}
\end{center}
\caption{\textbf{Device-independent tripartite QKD}.
(a) Tripartite QKD based on a triangle network. $a$, $b$, and $c$ ($x, y$, and $z$) denote the output (input) of $\textsf{A}_1$, $\textsf{A}_2$ and $\textsf{A}_3$, respectively. (b) Device-independent tripartite QKD. The output $e$ of eavesdropper $\textsf{E}$ depends on its input $E$ and these shared systems.}
\label{fig3}
\end{figure}

\textbf{Device-independent multipartite quantum key distribution}. Quantum entanglement is useful for distributing random key for secure communication \cite{5,6,7,8}. Our goal is to present a device-independent QKD with multiple parties using quantum networks. For simplicity, consider a tripartite network consisting of $\textsf{A}_1$, $\textsf{A}_2$ and $\textsf{A}_3$ (Fig.\ref{fig3}(a)). Let $x, y$, and $z$, respectively, denote the inputs of the three parties, and $a, b$, and $c$ denote the corresponding outcomes. $\rho_{ABC}$ denotes the total system. In a device-independent multipartite QKD (DIMQKD), the information available to an eavesdropper Eve is represented by $E$ correlated with $\rho_{ABC}$ (Fig.\ref{fig3}(b)). The total state is then denoted by $\rho_{ABCE}$ which should satisfy $\textrm{tr}_{E}\rho_{ABCE}=\rho_{ABC}$. Here, suppose that Eve wants to learn the output $a$ of $\textsf{A}_1$. For one copy resource, the predictability for Eve is quantified by the guessing probability of $a$ as $P_{gs}(a)=\max_aP(a|x)$, which is bounded by \cite{SI}:
\begin{eqnarray}
P_{gs}(a)\leq \frac{1}{2}+\frac{1}{12}\sqrt{72-\varpi^2_{es}}
\label{eq13}
\end{eqnarray}
where $\varpi_{es}$ denotes the estimate of the inequality (\ref{eq5}) using the practical quantum correlations in accordance with the practical quantum network (Fig.\ref{fig3}(a)). For $s$ copies of resources,the asymptotic secret-key rate $R:=-\frac{\log_2P_{gs}({\bf a}|E)}{s}$ \cite{20,38,52} is given by \cite{SI}:
\begin{eqnarray}
R\geq -\log_2(\frac{1}{2}+\frac{1}{12}
\sqrt{72-\varpi_{es}^2})-H(a|b,c)
\label{eq15}
\end{eqnarray}
where $P_{gs}({\bf a}|E)$ denotes the expected optimal guessing probability conditional on the eavesdropper's input $E$. The first term of Eq.(\ref{eq15}) represents the knowledge of Eve for the privacy amplification, and $H(a|b,c)$ quantifies the information needed for the error correction by $\textsf{A}_1$. Similar evaluations may be conducted for other networks.

\begin{figure}
\begin{center}
\resizebox{240pt}{240pt}{\includegraphics{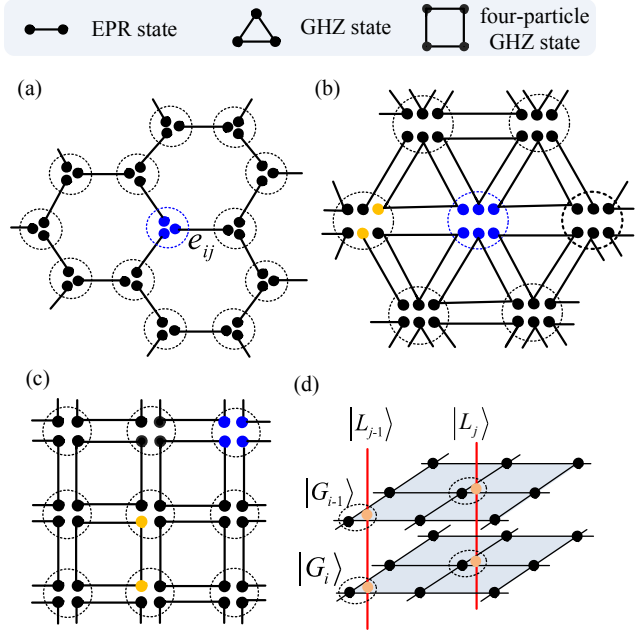}}
\end{center}
\caption{\textbf{Quantum networks for measurement-based quantum computation}.(a) Honeycomb network consisting of singlets $e_{ij}$'s  \cite{40}. (b) Triangular network consisting of GHZ states. (c) Square network consisting of four-particle GHZ states. (d) Cyclic network consisting of 2-dimensional graph states $|G_i\rangle$'s and 1-dimensional graph states $|L_j\rangle$'s.}
\label{fig4}
\end{figure}

\begin{figure}
\begin{center}
\resizebox{220pt}{110pt}{\includegraphics{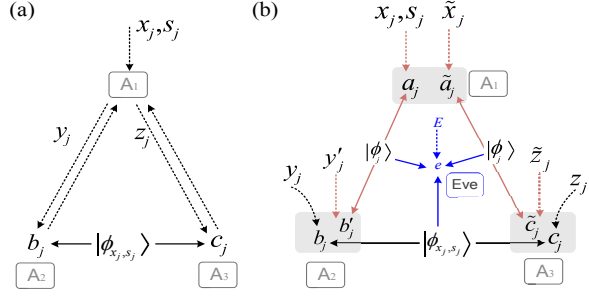}}
\end{center}
\caption{\textbf{DIBQC with two servers}.(a) A two-server BQC scheme. $\textsf{A}_1$ delegates a computational task to two servers $\textsf{A}_2$ and $\textsf{A}_3$ who share some EPR states of $\otimes_j|\phi_{x_j,s_j}\rangle$. $y_j$ and $z_j$ are bit encodings of the respective $\theta_j'$ and $\theta_j'+b_j\pi$. (b) DIBQC with two servers.  $\textsf{A}_1$ and $\textsf{A}_2$ (or $\textsf{A}_3$) share a random key $k_j$ (or $\hat{k}_j$) using the shared EPR states (QKD with purple lines).  The measurement outcomes $y_j,z_j,b_j$, and $c_j$ are then encrypted with the one-time pad.  An eavesdropper $\textsf{E}$ can correlate its system with the shared sources.}
\label{fig5}
\end{figure}

\textbf{Device-independent blind quantum computation}. Measurement-based quantum computation (MQC) provides a new computation model that makes use of simple measurement on qubits prepared in a highly entangled state \cite{40}. Some examples are shown in Fig.\ref{fig4} by using specific systems \cite{41,42}. The inequality (\ref{eq5}) provides a useful way to verify these computational resources. Another way is assisted with LOCC \cite{SI}. MQC allows one party with limited computational power to use other parties' computational resources without revealing the real task, inputs and outcomes \cite{43,44,45}. So-called blind quantum computation (BQC) provides a secure way to verify adversarial quantum processor utilizing the statistics of local measurements \cite{44,47}. Inspired by a recent scheme \cite{48}, a two-server BQC scheme is shown in Fig.\ref{fig5}(a) and DIBQC scheme is shown as (Fig.\ref{fig5}(b)) \cite{SI}:
\begin{itemize}
\item[(i)] $\textsf{A}_1$ (verifier) secretly chooses independent strings $\mathbf{x}=x_1\cdots{} x_n$ and $\mathbf{s}=s_1\cdots{} s_n$ with the uniform probability distribution.
\item[(ii)] $\textsf{A}_2$ and $\textsf{A}_3$ share some EPR states:  $\otimes_{j=1}^m|\phi_{x_j,s_j}\rangle$, where $|\phi_{x_j,s_j}\rangle=\frac{1}{\sqrt{2}}(
    \mathbbm{1}\otimes\sigma_x^{s_j}
    \sigma_z^{x_j})(|00\rangle+|11\rangle)$ with Pauli operators $\sigma_x$ and $\sigma_z$.
\item[(iii)]$\textsf{A}_1$ shares some EPR states $\otimes_{j}|\phi_{j}\rangle$ with $\textsf{A}_2$ and $\textsf{A}_3$, respectively.
\item[(iv)] $\textsf{A}_1$ measures the $j$-th entanglement shared with $\textsf{A}_2$ in the basis $\{\frac{1}{\sqrt{2}}
    (|0\rangle\pm{}e^{-i\theta_j'}|1\rangle)\}$ after a local qubit operation $U=|0\rangle\langle0|+e^{-2i\theta_j'}|1\rangle\langle 1|$, in which $\theta_j'=(-1)^{s_j}\theta_j+x_j\pi$ and $\theta_j$ is randomly chosen from $\{\frac{k\pi}{4}|k=0, 1, \cdots, 7\}$. $\textsf{A}_2$ measures the $j$-th entanglement shared with $\textsf{A}_1$ in the basis randomly chosen from $\{\{\frac{1}{\sqrt{2}}
    (|0\rangle\pm{}e^{-i\theta_j'}|1\rangle)\}, \forall j\}$. They detect eavesdroppers with public discussions, and then distil a random key $k_1\cdots{}k_t$.
\item[(v)] $\textsf{A}_2$ measures the local qubit of the $j$-th Bell state shared with $\textsf{A}_3$ in the basis $\{\frac{1}{\sqrt{2}}
    (|0\rangle\pm{}e^{-i\theta_j'}|1\rangle)\}$, and then sends the outcome $b_j\in\{0,1\}$ in accord with the one-time pad with $k_1\cdots{}k_t$.
\item[(vi)] $\textsf{A}_1$ and $\textsf{A}_3$ start the modified single-server protocol \cite{43} with the following encoding:  $\hat{\theta}_j\mapsto\hat{\theta}'_j+b_j\pi$, $j=1, 2, \cdots, m$. Here, the modified scheme means that all of $\textsf{A}_1$'s classical information are encoded into the one-time pad using the key which is shared by $\textsf{A}_1$ and $\textsf{A}_3$ \cite{44}.
\end{itemize}

The main difference between two protocols in Fig.5 is the transmission of classical information. $\textsf{A}_1$ uses the statistics to verify the shared random key with $\textsf{A}_2$ or $\textsf{A}_3$. Hence, the present DIBQC is sensitive to cheating \cite{45,48}. $\textsf{A}_3$ cannot obtain $\textsf{A}_2$'s outputs from $\textsf{A}_1$'s inputs \cite{47} because two servers are forbidden to communicate in the present model. Similar to single-server protocol \cite{45}, both servers cannot learn the hidden computational task. From Eq.(\ref{eq13}) the guessing probability about $\textsf{A}_1$'s outcome is given by $P_{gs}({\bf a}|E)\leq (\frac{1}{2}+\frac{1}{12}
\sqrt{72-\varpi_{es}^2}+m^{-1/4}_{es})^m$, where $m_{es}$ denotes the number of distilled key. One may further estimate the guessing probability for $\textsf{A}_1$'s computational task \cite{44,45}. Here, $\textsf{A}_1$ is able to perform quantum measurement while EPR states should be prepared and distributed by the servers \cite{45,48}.

\textbf{Device-independent quantum secret sharing}.  In an $(n,k)$ secret sharing scheme, a secret $x$ is split into $n$ parts $x_1, x_2, \cdots, x_n$ such that $x$ is recoverable from any $k$ parts \cite{49,50}. Inspired by recent scheme with GHZ state \cite{MBNC}, our goal is to propose a device-independent quantum secret sharing (DIQSS) using quantum networks based on the inequality (\ref{eq5}). For simplicity, a $(3,2)$ DIQSS is constructed using the triangular network ${\cal N}_q$ in Fig.\ref{fig3}(a). For a secret bit $s$, the goal is to distribute the split parts $s_1, s_2$, and $s_3$ to $\textsf{A}_1$, $\textsf{A}_2$ and $\textsf{A}_3$, respectively, where $s_i\oplus{}s_j=s$ with $1\leq i\not=j\leq 3$. In Bell model (Fig.\ref{fig3}), supposes that $a_1=s_1$, $a_2=s_2$ and $b_2=s_3$ for specified inputs $x_i$, $y_i$, and $z_i$. This implies that any two parties can cooperate to recover the secret $s$, but one party fails. The violation of Mermin inequality \cite{32} can rule out the outer eavesdropper but not inner untrustful party. The main reason is that a local correlating operation of receiver is equivalent to a measurement dependent hidden variable model \cite{MBNC}. Suppose that $\textsf{A}_i$ is eavesdropper who recovers the output $a_1$ of $\textsf{A}_1$ with $i\not=1$. The guessing probability $P(e=a_1|\textbf{x},u)$ by eavesdroppers can be reshaped using the conditional probability $P(\textbf{a},e|\textbf{x},u)$ into a linear optimization as
\begin{eqnarray}
\!\!\!\!\!\!\max_{P(\textbf{a},e|\textbf{x},u)}
&&\!\!\!\!\!\!P(e=a_1|\textbf{x},u)
\label{eq52}
\\
\textrm{s.t.} &&\!\!\!\!\sum_{\textbf{a},e}P(\textbf{a},e|
\textbf{x},u)=1, \forall \textbf{x}, u
\label{eq53}
\\
&&\!\!\!\!\!\! P(\textbf{a},e|\textbf{x},u)\geq 0, \forall \textbf{a}, \textbf{x}, e, u
\label{eq54}
\\
&&\!\!\!\!\!\!\sum_{w}(P(\textbf{a},e|\textbf{x},u)
-P(\textbf{a},e|\textbf{x}^{\zeta},u))=0, \forall \zeta
\label{eq55}
\\
&&\!\!\!\!\!\! P(\textbf{a}|\textbf{x})=
\sum_{u,e}p(u)P(\textbf{a},e|\textbf{x},u)
\label{eq56}
\\
&&\!\!\!\!\!\!\sum_{i=1}^{\lfloor\frac{n}{2}\rfloor}{\cal L}_i(P(\textbf{a}|\textbf{x}))> 2^{n}\sqrt{2}-4\sqrt{2}+2
\label{eq57}
\end{eqnarray}
where $\textbf{a}=a_1\cdots a_n$, $\zeta{}$ is a subset of $\{a_1, \cdots, a_n\}$, and $\textbf{x}^{\zeta{}}=x_1^\zeta{}\cdots x_n^\zeta{}$ satisfies $x_i^\zeta{}=x_i$ with $a_i\not\in{}\zeta{}$. Eq.(\ref{eq55}) is non-signalling condition for any subset $\zeta{}$ with respect to different inputs of the corresponding parties. Eqs.(\ref{eq56}) and (\ref{eq57}) are key to estimate the guessing probability of $P(e=a_1|\textbf{x},u)$, where the projected distribution $P(\textbf{a}|\textbf{x})$ is the $n$-variable probability derived from local measurements on ${\cal N}_q$. The linear optimization problem is algorithmically solvable using the semi-definite program \cite{54} and NPA hiearchy \cite{51,MBNC,SI}. Eve's guess probability $p(e=c|x_1,x_2,x_3)$ in quantum settings satisfies $p(e=c|x_1,x_2,x_3)\leq 3/4$ for $\varpi\geq 8.25$ and $p(e=c|x_1,x_2,x_3)=1/2$ for $\varpi=6\sqrt{2}$. This provides a useful method for characterizing secure quantum secret sharing. Similar optimization problems can be formed for parties being eavesdroppers.

\subsection*{Conclusion}

The multipartite correlations generated by a $k$-independent network form a star-convex set \cite{18}. There is a convex correlation set with relaxed input assumption arising from the inequality (\ref{eq5}). Compared with previous nonlinear Bell inequalities \cite{15,16,17,18}, the inequality (\ref{eq5}) enables a standard Bell test without assumptions of source independence for verifying general quantum networks, including acyclic networks \cite{18}, cyclic networks \cite{24}, and universal computational resources for measurement-based quantum computation \cite{41,42,43}. The main drawback is the lower visibility for noisy networks \cite{SI}. It may be applicable when both the number of parties and the number of entangled sources are not large enough. This remains an interesting problem for further improvement.

The independence of sources cannot be guaranteed in network scenarios because any untrusted party may correlate the shared independent sources into a new one. This is essentially different from the independence assumption of measurement settings \cite{55}, which is resolvable both in theory \cite{56} and experimental improvements \cite{57,58}. The present inequality (\ref{eq5}) allows remote legitimate parties to share a common random key in a fully device-independent manner. Another example is device-independent BQC \cite{45,47}. The last example is sharing of a classical secret \cite{51}. The formal security proofs of these applications and other candidates should be of interest in quantum information processing using practical quantum resources.

\section*{Acknowledgements}

We thank the helps of Ronald de Wolf, Carlos Palazuelos, Luming Duan, Yaoyun Shi, M. Orgun, J. Pieprzyk, Yuan Su, Huiming Li, Xiubo Chen, Yixian Yang. This work was supported by the National Natural Science Foundation of China (No.61772437), Sichuan Youth Science and Technique Foundation (No.2017JQ0048), Fundamental Research Funds for the Central Universities (No.2018GF07), and EU ICT COST CryptoAction (No.IC1306).

\newpage

\appendix
\section{Proof of the inequality (1)}

Consider an $n$-partite quantum network ${\cal N}_q$ (as shown in Fig.1(a) in the main text) consisting of $n$ parties $\textsf{A}_1,\cdots, \textsf{A}_n$. Suppose that the total state of ${\cal N}_q$ is in a biseparable state $\rho$ on Hilbert space $\otimes_{i=1}^n{\cal H}_{i}$ \cite{4}, that is, it can be decomposed into
\begin{eqnarray}
\rho=\sum_{I}p_{I}\varrho_{I,\overline{I}}
\label{bs}
\end{eqnarray}
where $I=\{s_1,\cdots, s_k\}\subset \{1, \cdots, n\}$ and $\overline{I}=\{j|j\not\in I, 1\leq j\leq n\}$ are bipartition of $\{1, \cdots, n\}$, $\{p_{I}\}$ is a probability distribution, $\varrho_{I,\overline{I}}$ denotes separable state on Hilbert space ${\cal H}_{I}\otimes {\cal H}_{\overline{I}}$, ${\cal H}_{I}=\otimes_{j\in I}{\cal H}_j$ and ${\cal H}_{\overline{I}}=\otimes_{j\in \overline{I}}{\cal H}_j$. Here, the summation is over all the bipartitions of $\{1, \cdots, n\}$.

Denote $M_{x_i}^{a_i}$ as positive-operator-value measurement (POVM for short) performed by $\textsf{A}_{i}$, where $M^{a_i}_{x_i}$ satisfies $\sum_{x_i=0,1}M^{a_i}_{x_i}=\mathbbm{1}$ for $a_i\in \{-1,1\}$, $i=1, \cdots, n$. These measurement operators satisfy the commutativity condition $[M^{a_i}_{x_i},M^{a_j}_{x_j}]=M^{a_i}_{x_i}M^{a_j}_{x_j}
-M^{a_j}_{x_j}M^{a_i}_{x_i}=0$ for any $i$ and $j$ with $i\not=j$. The binary output $a_i\in \{-1,1\}$ is conditional on the input $x_i\in\{0,1\}$. All parties perform local measurements on the state $\rho$. From Born rule, the output statistics is characterized by the joint probability conditional on inputs, that is
\begin{eqnarray}
P(\textbf{a}|\textbf{x})={\rm tr}(M_{x_{1}}^{a_1}\otimes\cdots{} \otimes{}M_{x_{n}}^{a_n}\rho)
\end{eqnarray}
where $\textbf{x}=x_1\cdots{}x_n$, and $\textbf{a}=a_1\cdots{}a_n$. For each input $x_i$, define $M_{x_i}$ as a dichotomic observable given by $M_{x_i}=M_{x_i}^{a_i=0}-M_{x_i}^{a_i=1}$, $i=1, \cdots, n$. With the distribution of $P(\textbf{a}|\textbf{x})$, an $n$-partite correlation in terms of $\rho$ is defined by
\begin{eqnarray}
\langle M_{x_{1}}\cdots{} M_{x_{n}}\rangle&=&{\rm tr}(M_{x_{1}}\otimes\cdots{} \otimes{}M_{x_{n}}\rho)
\nonumber\\
&=&\sum_{\textbf{a}}(-1)^{\sum_{i=1}^n(a_{i}+3)/2}
P(\textbf{a}|\textbf{x})
\label{A2}
\end{eqnarray}

In what follows, each party makes use of multiple observables. Especially, for each bipartition $I$ and $\overline{I}$, define $M_{x_j(I)}$ as dichotomic observable of $\textsf{A}_j$ for the input $x_j$ depending on $I$, that is, the input $x_j$ depends on $I$. For simplicity, denote
\begin{eqnarray}
&\textbf{M}_{I}=\prod_{j\in I}M_{x_j(I)}, \textbf{M}_{\overline{I}}=\prod_{j\in \overline{I}}M_{x_j(\overline{I})}
\label{A3}
\end{eqnarray}

Since the left side of the inequality (2) is linear for all the measurement observables $M_{x_{j}}$. All the biseparable states consist a convex set of $\{\sum_{I,\overline{I}}p_{I,\overline{I}}\varrho_{I}\otimes \varrho_{\overline{I}}\}$. So, the maximal bound of $\sum_{i=1}^{\lfloor\frac{n}{2}\rfloor}{\cal L}_i$ is achievable at the vertex of $\{\sum_{I,\overline{I}}p_{I,\overline{I}}\varrho_{I}\otimes \varrho_{\overline{I}}\}$, that is, product states. Hence, it is sufficient to prove the inequality (2) by using the product state $\rho=\varrho_{I}\otimes \varrho_{\overline{I}}$.

From the inequality (2) $CH_{I}$ is defined by
\begin{eqnarray}
CH_{I}=(\textbf{M}_{I}+\hat{\textbf{M}}_{I})
\textbf{M}_{\overline{I}}
+(\textbf{M}_{I}-\hat{\textbf{M}}_{I})
\hat{\textbf{M}}_{\overline{I}}
\label{A-1}
\end{eqnarray}
which is multipartite CHSH-type operator \cite{31} or Mermin operator \cite{40}, where
$|\textbf{M}_{I}|, |\hat{\textbf{M}}_{I}|, |\textbf{M}_{\overline{I}}|, |\hat{\textbf{M}}_{\overline{I}}|\leq 1 $.

Now, for each bipartition $I_0$ and $\overline{I}_0$ (which may be different from the bipartition $I$ and $\overline{I}$) there are two subcases for any product state $\rho=\varrho_{I}\otimes \varrho_{\overline{I}}$.
\begin{itemize}
\item[(i)] The state $\rho$ is separable in terms of the bipartition $I_0$ and $\overline{I}_0$. In this case, we can regard all the parties $\textsf{A}_{j}$ with $j\in I$ as one party, while all the parties $\textsf{A}_{t}$ with $t\in \overline{I}$ as the other. The quantity of $CH_{I}$ in Eq.(\ref{A-1}) is then regarded as a bipartite CHSH-type quantity \cite{31}. It follows that
    \begin{eqnarray}
    CH_{I_0}\leq 2
    \label{A4}
    \end{eqnarray}
   which holds for any product state $\rho=\varrho_{I}\otimes \varrho_{\overline{I}}$.
\item[(ii)] The state $\rho$ is entangled in terms of the bipartition $I_0$ and $\overline{I}_0$. In this case, there are two parties $\textsf{A}_{s}$ and $\textsf{A}_{t}$ with $s, t\in I$, or $\overline{I}$ such that they are entangled in the bipartition $I_0$ and $\overline{I}_0$, and $s\in I_0, t\in \overline{I}_0$. It follows that
   \begin{eqnarray}
   CH_{I_0}\leq 2\sqrt{2}
   \label{A5}
   \end{eqnarray}
  Here, the upper bound $2\sqrt{2}$ may be achievable. Take $\varrho_{I}=|\phi\rangle_{12}\langle\phi|$ and $\varrho_{\overline{I}}=|\phi\rangle_{34}\langle\phi|$ as an example, where $|\phi\rangle$ is an EPR state. $I_0=\{1,3\}$ and $\overline{I}_0=\{2,4\}$. $\textsf{A}_{1}$  and $\textsf{A}_{2}$ shares one EPR state. The maximal bound in the inequality (\ref{A5}) can be followed from the CHSH test \cite{31}.
\end{itemize}

Moreover, for each product state of $\rho=\varrho_{I}\otimes \varrho_{\overline{I}}$ with the given bipartition $I$ and $\overline{I}$, there is at least one term $CH_{I}$ in the summation of $\sum_{i=1}^{\lfloor{\frac{n}{2}}\rfloor}{\cal L}_i$ satisfies that  $CH_{I}\leq 2$. Combined with the linearity of ${\cal L}_i$, it follows that
\begin{eqnarray}
\sum_{i=1}^{\lfloor{\frac{n}{2}}\rfloor}{\cal L}_i\leq 2(N-1)\sqrt{2}+2
\label{A6}
\end{eqnarray}
for any product state $\rho$, where $N$ denotes the total number of $CH_{I}$s.

In what follows, we need to evaluate $N$. For an even integer $n$, define $m_i$ as the total number of $CH_{I}$ for a given $i$ with $i\leq \frac{n}{2}$. From the standard combination theory \cite{Graph} we obtain
\begin{eqnarray}
m_i&=&C(n,i), 1\leq i<\frac{n}{2},
\nonumber
\\
m_j&=&\frac{1}{2}C(n,j), j=\frac{n}{2}
\label{A7}
\end{eqnarray}
where $C(n,i)$ denotes the combination number of choosing $i$ balls from $n$ different balls without choosing order. From Eq.(\ref{A3}), it follows that
\begin{eqnarray}
N&=&\sum_{i=1}^{n/2}m_i
\nonumber\\
&=&\frac{1}{2}\sum_{i=1}^{n-1}C(n,i)
\label{A8}
\\
&=&\frac{1}{2}\sum_{i=0}^{n}C(n,i)-1
\label{A9}
\\
&=&2^{n-1}-1
\label{A10}
\end{eqnarray}
where Eq.(\ref{A8}) is obtained by using the equalities: $C(n,i)=C(n,n-i)$ with $i=1, \cdots, \frac{n}{2}-1$. Eq.(\ref{A9}) is from the equalities: $C(n,0)=C(n,n)=1$.  Eq.(\ref{A10}) is from the equality: $\sum_{i=0}^nC(n,i)=2^n$.

For an odd integer $n$, similar to Eq.(\ref{A7}), using the standard combination theory we obtain
\begin{eqnarray}
m_i=C(n,i), 1\leq i\leq \lfloor{}\frac{n}{2}\rfloor
\label{A11}
\end{eqnarray}
where $\lfloor{}x{}\rfloor$ denotes the maximal integer which is no more than $x$. Similar to Eqs.(\ref{A7})-(\ref{A10}) it is easy to prove that
\begin{eqnarray}
N=&\sum_{i=1}^{\frac{n}{2}}m_i
=2^{n-1}-1
\label{A12}
\end{eqnarray}

From Eqs.(\ref{A6}) and (\ref{A10}), and the linearity of ${\cal L}_i$, it yields that
\begin{eqnarray}
\sum_{i=1}^{\lfloor{\frac{n}{2}}\rfloor}{\cal L}_i\leq 2^n\sqrt{2}-4\sqrt{2}+2
\label{A13}
\end{eqnarray}

For a general quantum network, $CH_{I}\leq 2\sqrt{2}$ for each bipartition $I$ and $\overline{I}$. From Eqs.(\ref{A6}), (\ref{A10}) and (\ref{A12}), it follows that
\begin{eqnarray}
\sum_{i=1}^{\lfloor{\frac{n}{2}}\rfloor}{\cal L}_i\leq (2^n-2)\sqrt{2}
\label{A14}
\end{eqnarray}
The upper bound is achievable for some quantum networks, which will be proved in the next section. This completes the proof.

\textbf{Remark S1}. The present inequality (2) may be trivial if all observables of $\textbf{M}_{I}, \hat{\textbf{M}}_{I}, \textbf{M}_{\overline{I}}, \hat{\textbf{M}}_{\overline{I}}$ are different for each bipartition $I$ and $\overline{I}$, where each party has $2$ inputs and two outputs for each bipartition $I$ and $\overline{I}$. In this case, the present inequality (2) is a summation of different CHSH inequalities \cite{31}, where each party has $2N$ inputs and two outputs. However, it is not trivial when each party has less than $2N$ inputs, or only a few inputs. This is the main concern in what follows. Generally, the number of inputs will depend on the involved state or quantum network. We can prove that the total number of inputs is no more than $8$. This will be followed from the proof for quantum networks in the next sections.

\textbf{Example S1}. Consider a triangle network ${\cal N}_q$ \cite{13} shared by three parties Alice, Bob and Charlie. We get the following inequality:
\begin{eqnarray}
&&A_1B_1C_1+A_2B_1C_1+A_1B_2C_2-A_2B_2C_2
\nonumber\\
&&+A_3B_3C_1+A_3B_4C_1+A_4B_3C_2-A_4B_4C_2
\nonumber\\
&&+A_3B_5C_3+A_3B_5C_4+A_4B_6C_3-A_4B_6C_4
\leq 4\sqrt{2}+2
\label{C1}
\end{eqnarray}
which holds for all biseparable states. $A_i, B_j$, and $C_k$ are dichotomic observable of $\textsf{A}_1$, $\textsf{A}_2$, and $\textsf{A}_3$, respectively, $i,k=1, \cdots, 4; j =1, \cdots, 6$. It means that Alice and Charlie have 4 inputs while Bob has 6 inputs.

\textbf{Example S2}. Different from Example 1, we can get another inequality with less inputs as:
\begin{eqnarray}
&&A_1B_1C_1+A_2B_1C_1+A_1B_2C_2-A_2B_2C_2
\nonumber\\
&&+A_3B_3C_1+A_3B_4C_1+A_4B_3C_2-A_4B_4C_2
\nonumber\\
&&+A_3B_1C_3+A_3B_1C_4+A_4B_2C_3-A_4B_2C_4
\leq 4\sqrt{2}+2
\label{C2}
\end{eqnarray}
which holds for all biseparable states. It means that each party has 4 inputs.

\section{Genuinely multipartite nonlocality of quantum networks with generalized EPR states}

Let ${\cal N}_q$ be an $n$-partite connected quantum network shared by $\textsf{A}_1, \cdots, \textsf{A}_n$. Suppose that ${\cal N}_q$  consists of generalized EPR states $|\phi_1\rangle, \cdots, |\phi_m\rangle$, where $|\phi_i\rangle$ is given by
\begin{eqnarray}
|\phi_i\rangle=\cos_i|00\rangle+\sin\theta_i|11\rangle
\label{B-1}
\end{eqnarray}
$\theta_i\in (0,\frac{\pi}{4})$. Here, the connectedness of quantum network means that for each pair of $\textsf{A}_i,\textsf{A}_j$, there is a chain-type subnetwork ${\cal N}_{i\to j}=\{\textsf{A}_i,\textsf{A}_{s_1}, \cdots, \textsf{A}_{s_k},\textsf{A}_j\}$, where two adjacent parties share entangled states. The main goal is to prove the genuinely multipartite nonlocality of ${\cal N}_q$ with even $n$ in this section. The case of odd $n$ will be proved in the next section.

\textbf{Result S1}. There are local observables for each party such that the output statistics violates the inequality (2) with an even integer $n\geq 4$, that is, ${\cal N}_q$ is genuinely multipartite nonlocal in the biseparable model.

\textbf{Lemma S1}. For an even $n\geq 4$, ${\cal N}_q$ can be locally transformed into a new network ${\cal N}_q'$ without classical communication such that all parties have odd particles, where ${\cal N}_q'$ consists of generalized EPR states and $4$-particle GHZ states.

\textbf{Proof of Lemma S1}. By representing an EPR state as one edge with two nodes, ${\cal N}_q$ can be schematically represented by a connected graph ${\cal G}$ in which each party owns some nodes of edges. Here, a connected graph means that for any two nodes, there is a path connecting them. For each pair of two parties who share EPR states are connected by edges. The result is easily followed for an acyclic graph which has no cycle, that is, a tree graph ${\cal G}$ \cite{Graph}. In fact, note that the total number of particles is even. The total number of parties who have odd number of particles is even. It means that the total number of parties who have even number of particles is even. Consider one pair $\textsf{A}_i$ and $\textsf{A}_j$ who has even number of particles. Since ${\cal G}$ is connected, there is a connected subgraph ${\cal G}_{i\to{}j}=\{\textsf{A}_i,\textsf{A}_{s_1}, \cdots, \textsf{A}_{s_k},\textsf{A}_j\}$, where $\textsf{A}_{s_1}, \cdots, \textsf{A}_{s_k}$ have odd number of particles. In this case, each adjacent pair of ${\cal G}_{i\to{}j}$ can transform one shared EPR state $|\phi_i\rangle|0\rangle|0\rangle$ into a four-particle GHZ state $|\hat{\phi}_i\rangle=\cos\theta_i|0000\rangle+\sin\theta_i|1111\rangle$ by using local controlled-not operation and one axillary particle in the state $|0\rangle$ for each party. After these local operations being performed, all the parties in ${\cal G}_{i\to{}j}$ have odd number of particles. This procedure can be iteratively completed for all parties who have even number of particles.

In what follows, we propose Algorithm 1 to locally transform a cyclic graph. Here, a cyclic graph ${\cal G}$ means that it contains at least one cycle.

\begin{algorithm}[h!]
\caption{Transforming ${\cal N}_q$ with local unitary operations and axillary particles in the state $|0\rangle$}
\KwIn{A connected graph ${\cal G}$ with $n$ vertexes, where $n$ is even integer}
\KwOut{A connected graph ${\cal G}'$ with $n$ vertexes, where each vertex has odd number of nodes}
\begin{itemize}
\item[S0] ({\bf Initialization}) Each cyclic graph ${\cal G}$ is decomposed into a set of cyclic subgraphs $\{{\cal G}_1, \cdots, {\cal G}_k\}$, where ${\cal G}_i$ denotes a subgraph with one cycle in which any two adjacent parties share some EPR states. This can be efficiently completed \footnote{For $i$ from $1$ to $n$, find new cycle ${\cal G}_j$ started from the vertex $\textsf{A}_i$ such that ${\cal G}_j\in \{{\cal G}_1, \cdots, {\cal G}_{j-1}\}$. Otherwise, there is no new cycle. For each $j$, there are at least one new vertex which will be added in ${\cal G}_j$ going beyond ${\cal G}_1, \cdots, {\cal G}_{j-1}$. This procedure will stop if all the vertex will be included in some cycle. The total time complexity is $O(n^2)$, where each cycle has at most $n$ vertexes.}. Denote $\textsf{A}_i$s as the vertexes (or parties in the network) who have even number of nodes (or particles in the network). Denote $\textsf{B}_j$s as the vertexes (or parties in the network) who have odd number of nodes (or particles in the network), where the subindexes $i,j$ of $\textsf{A}_i$s and $\textsf{B}_j$s are different for convenience.
\item[S1]If all vertexes have odd number of nodes, then output the graph ${\cal G}$;

\item[S2]There are vertexes which have even number of nodes. For convenience, assume that in each subgraph ${\cal G}_i$ there are vertexes which have even number of nodes. Otherwise, delete the subgraphs without vertexes which have even number of nodes.
\begin{itemize}
\item[] For $\ell$ from $1$ to $k$, consider ${\cal G}_\ell$ as follows.
\begin{itemize}
\item[1)] For two adjacent vertexes $\textsf{A}_i$ and $\textsf{A}_{j}$ in ${\cal G}_\ell$, if both of them have even number of nodes in ${\cal G}_\ell$, they can transform one connected edge into one hyper edge with four nodes by adding one new node for each vertex \cite{Graph}. This can be completed by transforming one shared EPR state $|\phi_i\rangle|0\rangle_{i}|0\rangle_{j}$ into one four-particle GHZ state $|\hat{\phi}_i\rangle=\cos\theta_i|0000\rangle+\sin\theta_i|1111\rangle$ by using local controlled-not operations and one axillary particle in networks. After these operations being performed, both vertexes have odd number of nodes. They are then relabeled as $\textsf{B}_i$ and $\textsf{B}_{j}$.
\item[2)] For two adjacent vertexes $\textsf{A}_i$ and $\textsf{A}_{j}$ in ${\cal G}_\ell$, if one of them, $\textsf{A}_i$ for example, has even number of nodes in ${\cal G}_\ell$, there exist other vertexes $\textsf{A}_{s_1}, \cdots$, $\textsf{A}_{s_t}$ which own even number of nodes. Choose one vertex $\textsf{A}_{s_1}$ such that there are other vertexes $\textsf{B}_{j_1}, \cdots, \textsf{B}_{j_\ell}$ satisfying that $\textsf{A}_i, \textsf{B}_{j_1}, \cdots, \textsf{B}_{j_\ell}, \textsf{A}_{s_1}$ consist of a chain-shaped connected subgraph. Now, each pair of adjacent vertexes can transform one edge into one hyper edge with four nodes by adding one new node for each vertex. After the local operations being performed, these vertexes have odd number of nodes, where one node is added for $\textsf{A}_{i}$ and $\textsf{A}_{s_1}$, while two nodes are added for $\textsf{B}_{j_1}, \cdots, \textsf{B}_{j_\ell}$. $\textsf{A}_{i}$ and $\textsf{A}_{s_1}$ are then relabeled as $\textsf{B}_{i}$ and $\textsf{B}_{s_1}$ respectively. These two steps can be iteratively performed for all vertexes in ${\cal G}_\ell$.
\end{itemize}
\end{itemize}
\end{itemize}
\end{algorithm}

\begin{figure}
\begin{center}
\resizebox{300pt}{270pt}{\includegraphics{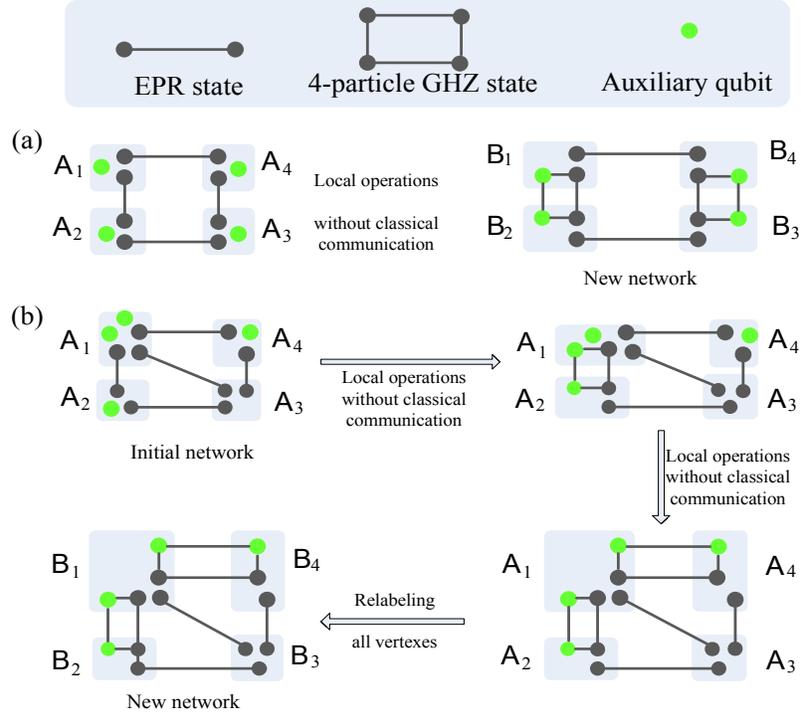}}
\end{center}
\caption{{\bf Schematically transforming a network using local operations and auxiliary particles without classical communication}. (a) Cyclic network with four parties. Four parties $\textsf{A}_1, \cdots, \textsf{A}_4$ share four EPR states. (b) Cyclic network with two cycles. Four parties $\textsf{A}_1, \cdots, \textsf{A}_4$ share five EPR states.}
\label{figs1}
\end{figure}

For ${\cal G}_\ell$, each vertex has only added even number of nodes, even if local transformations may change the number of nodes for these vertexes contained in ${\cal G}_j$ for $j<\ell$. In this case, all the vertexes $\textsf{B}_s\in{\cal G}_j$ have odd number of nodes after local transformations being performed by all vertexes in ${\cal G}_\ell$. It means that one does not need to consider all the subgraphs ${\cal G}_j$ with $j<\ell$ after ${\cal G}_\ell$ has been changed. This is key to stop Algorithm 1.

We take use of this algorithm to prove Lemma S1. Note that $n$ is even and each edge has two nodes. Moreover, two nodes will be added in each time. For each round, there are two vertexes $\textsf{A}_{i}$ and $\textsf{A}_{j}$ being relabeled as $\textsf{B}_{i}$ and $\textsf{B}_{j}$ while other vertexes $\textsf{B}_{s}$ are unchanged. By induction on the number $n$ of vertexes, all the vertexes $\textsf{A}_{j}$ can be relabeled as $\textsf{B}_{j}$ because the network ${\cal N}_q$ is connected. This can be proved by contradiction. In fact, assume that one vertex has even number of particles while $n-1$ vertexes have odd number of nodes. There are odd nodes because $n-1$ is odd integer. This contradicts with the fact that there is even number of vertexes from two nodes of each edge. This completes the proof of Lemma S1. $\Box$

\textbf{Example S3}. Two networks are shown in Fig.S\ref{figs1}. In Fig.S\ref{figs1}(a), there are four parties who share four EPR states. In step S2, $\textsf{A}_1$ and $\textsf{A}_2$ can locally change an EPR state into four-particle GHZ state with local operations and axillary particles. Similar result holds for $\textsf{A}_3$ and $\textsf{A}_4$. After these local operations being performed, they obtain a new network consisting of two EPR states and two four-particle GHZ states, where each party has three particles. In Fig.S\ref{figs1}(b), there are four parties who share five EPR states. Different from the network shown in Fig.S\ref{figs1}(a), there are two cycles, that is, ${\cal G}_1=\{\textsf{A}_1, \textsf{A}_2, \textsf{A}_3\}$ and ${\cal G}_2=\{\textsf{A}_1, \textsf{A}_3, \textsf{A}_4\}$. In step S2, consider ${\cal G}_1$ firstly. $\textsf{A}_1$ and $\textsf{A}_2$ can change an EPR state into four-particle GHZ state with local operations and axillary particles. And then, consider ${\cal G}_2$. $\textsf{A}_1$ has four particles by adding the particles involved in cycle ${\cal G}_1$. So, it is sufficient to consider the EPR state shared by $\textsf{A}_1$. Specifically, $\textsf{A}_1$ and $\textsf{A}_4$ transform the shared EPR state into four-particle GHZ state with local operations and axillary particles. After these local operations, they obtain a new network consisting of three EPR states and two four-particle GHZ states, where each party has odd number of particles.

\textbf{Proof of Result S1}. Let $ M_{x_i}$ be dichotomic observable of $\textsf{A}_{i}$ for the input $x_i\in\{0, 1\}$. In what follows, it is sufficient to prove that there are local observable for each party on ${\cal N}_q$ such that the quantum correlations generated by local measurements violate the inequality (2).

From Lemma S1, after local transformations being performed, each party $\textsf{A}_j$ is relabeled as $\textsf{B}_j$ who has $n_j$ number of particles. From Eq.(\ref{B-1}), the total state of ${\cal N}_q$ is given by
\begin{eqnarray}
\rho=\otimes_{i=1}^m|\phi_i\rangle\langle \phi_i|
\label{B2}
\end{eqnarray}
where $|\phi_i\rangle$ is EPR state or 4-particle GHZ state. Define $\varpi_q$ as the quantum violation given by
\begin{eqnarray}
\varpi_q=\sum_{i=1}^{\lfloor{\frac{n}{2}}\rfloor}{\cal L}_i
\label{B3}
%\label{eq17}
\end{eqnarray}
where ${\cal L}_i$s are defined in the inequality (2) associated with the state $\rho$ in Eq.(\ref{B2}). It is sufficient to evaluate $\varpi_q$ by maximizing all the multipartite quantities $CH_{I}$ in the inequality (2).

For each set $I=\{s_1, \cdots, s_k\}$, consider $CH_{I}$ firstly. Note that ${\cal N}_q$ is connected. For each pair of parties $\textsf{B}_{i}$ and $\textsf{B}_{j}$ with $i\in I$ and $j\not\in I$, there is one chain-shaped subnetwork ${\cal N}_{i\to j}\subset {\cal N}'_q$ connecting them. There is one longest chain-shaped subnetwork in these subnetworks $\{{\cal N}_{s_t\to \ell}\}$, that is, ${\cal N}_{s_t\to \ell}\subset {\cal N}'_q$ connecting the parties $\textsf{B}_{s_t}$ and $\textsf{B}_{\ell}$, where $s_t\in {}I$ and $\ell\not\in{}I$. For simplicity, define ${\cal N}_{s_t\to \ell}$ as
\begin{eqnarray}
{\cal N}_{s_t\to \ell}=\{\textsf{B}_{s_t},\textsf{B}_{l_1}, \cdots, \textsf{B}_{l_v}, \textsf{B}_{\ell}\}
\label{Subnetwork}
\end{eqnarray}
where $l_1, \cdots, l_v \not\in I$.

In what follows, define observables for each party as
\begin{eqnarray}
&&M_{x_{s_t}(I)}=\cos\theta\sigma_z^{\otimes{}n_{s_t}}
+\sin\theta\sigma_x^{\otimes{}n_{s_t}},
\label{B4}
\\
&&\hat{M}_{x_{s_t}(I)}=\cos\theta\sigma_z^{\otimes{}n_{s_t}}
-\sin\theta\sigma_x^{\otimes{}n_{s_t}},
\label{B5}
\\
&&M_{x_{j}(I)}=\sigma_z^{\otimes{} n_{j}}, \hat{M}_{x_{j}(I)}=\sigma_x^{\otimes{}n_j}, \forall j\not =s_t
\label{B52}
\end{eqnarray}
where $n_{j}$ denotes the number of particles owned by $\textsf{B}_{j}$, and $X^{\otimes k}$ denotes the tensor of $k$ copies of $X$. It is easy to show that $M_{s_1}M_{s_1}^\dag=\hat{M}_{s_1}\hat{M}_{s_1}^\dag=\mathbbm{1}$ because $n_1$ is odd integer. Here, $x_{j}(I)$ means that the input $x_j$ depends on the set $I$. With these observables, from Eqs.(\ref{A3}), (\ref{B4})-(\ref{B52}) it follows that
\begin{eqnarray}
\langle \textbf{M}_{I}\textbf{M}_{\overline{I}}\rangle&=&{\rm tr}((\cos\theta\sigma_z^{\otimes{}n_{s_t}}
+\sin\theta\sigma_x^{\otimes{}n_{s_t}})(\otimes_{j\not=s_t}\sigma_z^{\otimes{} n_{j}})\rho)
\nonumber
\\
&=&\cos\theta{\rm tr}((\sigma_z^{\otimes{}n_{s_t}}\otimes_{j\not=s_t}\sigma_z^{\otimes{} n_{j}})(\rho_{s_t}\otimes_{j\not=s_t} \rho_{j}))
\label{B5-4}
\\
&=&\cos\theta
\label{B5-5}
\end{eqnarray}
In Eq.(\ref{B5-4}), $\rho_{j}$ denotes the state owned by $\textsf{B}_{j}$, for $j\in \{1, \cdots, n\}$. Eq.(\ref{B5-4}) follows from Lemma S1. Note that ${\rm tr}(\sigma_x\otimes \sigma_z|\phi_i\rangle\langle \phi_i|)={\rm tr}(\sigma_z\otimes \sigma_x|\phi_i\rangle\langle \phi_i|)=0$ for any generalized EPR state $|\phi_i\rangle$. Moreover, ${\rm tr}(M_1\otimes M_2\otimes M_3\otimes M_4|\phi_i\rangle\langle \phi_i|)=0$ for any generalized 4-particle GHZ state $|\phi_i\rangle$ when $M_i\in \{\sigma_z,\sigma_x\}$ and $M_i$s are different. With these equalities, it follows from Eq.(\ref{B2}) that ${\rm tr}((\sigma_x^{\otimes{}n_{s_t}})(\otimes_{j\not=s_t}\sigma_z^{\otimes{} n_{j}})(\rho_{s_t}\otimes_{j\not=s_t} \rho_{j}))$. Moreover, we get Eq.(\ref{B5-5}) from the equalities of ${\rm tr}(\sigma_z\otimes \sigma_z|\phi_i\rangle\langle \phi_i|)=1$  for any generalized EPR state $|\phi_i\rangle$, and ${\rm tr}(\sigma_z^{\otimes 4}|\varphi_i\rangle\langle \varphi_i|)=1$ for any generalized 4-particle GHZ state $|\varphi_i\rangle$.

Similarly, we get that
\begin{eqnarray}
\langle\hat{\textbf{M}}_{I}\textbf{M}_{\overline{I}}\rangle&=&\cos\theta
\label{B61}\\
\langle\textbf{M}_{I}\hat{\textbf{M}}_{\overline{I}}\rangle&=&
\langle\hat{\textbf{M}}_{I}\hat{\textbf{M}}_{\overline{I}}\rangle=\sin\theta\prod_{j=1}^m\sin2\theta_j
\label{B63}
\end{eqnarray}
From Eqs.(\ref{A-1}) and (\ref{B5-5})-(\ref{B63}), it follows that
\begin{eqnarray}
CH_{I}&=&{\rm tr}(((\textbf{M}_{I}+\hat{\textbf{M}}_{I})
\textbf{M}_{\overline{I}}+(\textbf{M}_{I}-\hat{\textbf{M}}_{I})
\hat{\textbf{M}}_{\overline{I}})\rho)
\\
&=&2\cos\theta+2\sin\theta\prod_{j=1}^m\sin2\theta_j
\nonumber\\
&=&2\sqrt{1+\prod_{j=1}^m(\sin2\theta_j)^2}
\label{B6}
\end{eqnarray}
when $\theta$ satisfies $\cos\theta=1/\sqrt{1+\prod_{j=1}^m(\sin2\theta_j)^2}$.

Generally, for each set $I\subset\{1, \cdots, n\}$, from the assumption in the inequality (2), each party $\textsf{B}_{j}$ can choose proper observables depending on the input $x_{j}(I)$ such that the joint statistics satisfies the inequality (\ref{B6}). The main reason is that one may choose different input $x_{j}(I)$ depending on $I$. This implies that the inequality (\ref{B6}) holds for all the multipartite quantities $CH_{I}$s simultaneously. Hence, there exist observables for each observer such that the quantum correlations satisfy
\begin{eqnarray}
\varpi_q&=&\sum_{i}{\cal L}_i
\nonumber\\
&=&2N(1+\prod_{j=1}^m(\sin2\theta_j)^2)^{1/2}
\nonumber\\
&>&2^n\sqrt{2}-4\sqrt{2}+2
\label{B7}
\end{eqnarray}
if all $\theta_j$s satisfy the following inequality:
\begin{eqnarray}
\prod_{j=1}^m\sin2\theta_j
>(N^2-2(2-\sqrt{2})N+3-2\sqrt{2})^{1/2}
\label{B8}
\end{eqnarray}
where $N=2^{n-1}-1$. This completes the proof for even $n$. $\Box$

\section{Genuinely multipartite nonlocality of general connected quantum networks}

Let ${\cal N}_q$ be an $n$-partite connected quantum network consisted of $\textsf{A}_1, \cdots, \textsf{A}_n$. Suppose that ${\cal N}_q$ consists of generalized GHZ states $|\phi_1\rangle, \cdots, |\phi_m\rangle$, where $|\phi_i\rangle$ is given by
\begin{eqnarray}
|\phi_i\rangle=\cos_i|0\rangle^{\otimes m_i}+\sin\theta_i|1\rangle^{\otimes m_i}
\label{C-1}
\end{eqnarray}
$\theta_i\in (0,\frac{\pi}{4})$. The main goal in this section is to prove the genuinely multipartite nonlocality of ${\cal N}_q$ by using the inequality (2).

\textbf{Result S2}. There are local observables for each party such that the output statistics violates the inequality (2), that is, ${\cal N}_q$ is genuinely multipartite nonlocal in the biseparable model.

\textbf{Proof of Result S2}. Note that $m_i$ can be assumed to be even integer for each $i$. Otherwise, one party can change it into new GHZ state with even number of particles by using local operation and axillary particle. With this assumption, we firstly prove that ${\cal N}_q$ is genuinely $n$-partite nonlocal for even $n$ with $n\geq 3$. Similar to local transformations defined in Algorithm 1, ${\cal N}_q$ can be locally transformed into new network ${\cal N}'_q$ without classical communication such that all parties have odd number of particles, where ${\cal N}'_q$ consists of generalized EPR states and GHZ states with even number of particles. The proof is similar to Algorithm 1. The rest proof of the genuinely multipartite nonlocality is similar to the procedure from Eq.(\ref{B3}) to Eq.(\ref{B7}).

In what follows, we prove the result for odd $n$ with $n\geq 3$. Similar to local transformations defined in Algorithm 1, ${\cal N}_q$ can be locally transformed into a new network ${\cal N}'_q$ without classical communication such that there are $n-1$ observers who have odd number of particles, where ${\cal N}'_q$ consists of generalized EPR states and GHZ states with even number of particles. The proof is similar to Algorithm 1 by considering a connected subnetwork consisting of $n-1$ parties.

There are three subcases to prove the genuinely multipartite nonlocality of ${\cal N}'_q$.
\begin{itemize}
\item[S1] Consider the subnetwork consisting of $n-1$ parties $\textsf{A}_1, \cdots, \textsf{A}_{n-1}$. These parties can locally transform the shared network into another subnetwork using local unitary operations without the help of classical communication such that they have odd number of particles. After these local transformations, these parties are relabeled by $\textsf{B}_1, \cdots, \textsf{B}_{n-1}$. The rest proof is similar to Eqs.(\ref{B2})-(\ref{B7}) when $\textsf{A}_n$ has odd number of particles. Otherwise, there are two kinds of multipartite CHSH-type quantities $CH_{I}$ which will be distinguished from their inputs.
\item[S2]  Consider $CH_{I}$ with $n\not\in {}I$.  Similar to Eq.(\ref{Subnetwork}), there exists one longest chain-shaped network  ${\cal N}_{s_t\to \ell}=\{\textsf{B}_{s_t}, \textsf{B}_{j_1}, \cdots, \textsf{B}_{j_t}, \textsf{B}_{\ell}\}$, where $j_1, \cdots, j_t, \ell\not\in I$. For the party $\textsf{A}_n$, define its local observable as
\begin{eqnarray}
&&M_{x_n(I)}=\sigma_z^{\otimes{}h_n},
  \nonumber  \\
&&\hat{M}_{x_n(I)}=\sigma_x^{\otimes{}h_n}
\end{eqnarray}
where $h_n$ denotes the number of particles owned by the party $\textsf{B}_{n}$. With this definition, the inequality (\ref{B5}) holds for each $I\subset\{1, \cdots, n\}$ satisfying $n\not \in {}I$. From the assumption in the inequality (2) in the main text, i.e., $t_{I},\hat{t}_{I}$ are different for any $I$, the party $\textsf{A}_{n}$ can choose proper local observable for these post-selective terms $CH_{I}$ simultaneously. Hence, the inequality (\ref{B6}) holds for all $I$s satisfying $n\not \in{}I$.
\item[S3] Consider $CH_{I}$ with $n\in {}I$. In this case, from the assumption of the inequality (2), for each $I$ $\textsf{A}_n$ and $\textsf{A}_i$, who share at least one GHZ state can transform locally the shared states, such that $\textsf{A}_n$ owns odd number of particles after local operations. The observables of $\textsf{A}_n$ are similar to these given in Eqs.(\ref{B3}) and (\ref{B4}). And then, there exist another set of observables for all parties such that the inequality (\ref{B6}) holds for $CH_{I}$. Note that all $CH_{I}$s can be distinguished by all parties using proper inputs. This means that the inequality (\ref{B6}) holds for all the terms $CH_{I}$. Consequently, the inequality (\ref{B6}) holds for each odd $n$.
\end{itemize}
This completes the proof. $\Box$

\section{Evaluating the inputs of the inequality (2)}

In this section, we show that there are eight inputs for each party involved in the inequality (2) for verifying general networks. The main idea is from the proof in Appendixes B and C.

Consider an $n$-partite connected quantum network ${\cal N}_q$ consisting of EPR states and GHZ states. For an even $n\geq 3$, from Lemma S1 ${\cal N}_q$ can be locally transformed into new network ${\cal N}'_q$ satisfying that each party has odd number of particles, where ${\cal N}'_q$ consists of generalized EPR states and GHZ states with even number of particles.

Similar to Eq.(\ref{Subnetwork}), for each $I\subset\{1, \cdots, n\}$, there is one largest chain-type subnetwork ${\cal N}_{s_t\to \ell}=\{\textsf{B}_{s_t},\textsf{B}_{l_1}, \cdots, \textsf{B}_{l_v}, \textsf{B}_{\ell}\}$ connecting two parties $\textsf{B}_{s_t}$ and $\textsf{B}_{\ell}$ with $s_t\in I$ and $l_1, \cdots, l_v, \ell \not\in I$. The quantum network ${\cal N}'_q$ can be verified by using the inequality (2) with four inputs. In fact, for each party $\textsf{B}_j$, there are two subcases.
\begin{itemize}
\item[(i)]$j=s_t$. In this case, there are two input indexes $x_j(I)\in\{1, 2\}$ associated with the observables shown in Eqs.(\ref{B4}) and (\ref{B5}).
\item[(ii)] $j\not=s_t$. In this case, there are two input indexes $x_j(I)\in\{3, 4\}$ associated with the observables shown in Eq.(\ref{B52}).
\end{itemize}
Note that $s_t$ depends the longest chain-type subnetwork ${\cal N}_{s_t\to \ell}$, which may be relaxed as a general chain-type subnetwork without requirement of the length. With this relaxation, one can assume that $s_t=\min \{s_j|s_j\in I\}$ and $\ell=\min\{j|j\not\in I\}$.  Denote
\begin{eqnarray}
&&\textbf{M}_{I}=M_{x_{s_t}=1}\prod_{j\in I, j\not=s_t}M_{x_j=3},
\nonumber\\
&&\hat{\textbf{M}}_{I}=M_{x_{s_t}=2}\prod_{j\in I, j\not=s_t}M_{x_j=4},
\nonumber\\
&&\textbf{M}_{\overline{I}}=\prod_{j\in \overline{I}}M_{x_j=3},
\nonumber\\
&& \hat{\textbf{M}}_{\overline{I}}=\prod_{j\in \overline{I}}M_{x_j=4}
\label{BB}
\end{eqnarray}
With these definitions, the inequality (2) can be rewritten into
\begin{eqnarray}
\sum_{i=1}^{\lfloor\frac{n}{2}\rfloor}{\cal L}_i\leq 2^{n}\sqrt{2}-4\sqrt{2}+2
\label{eq5s}
\end{eqnarray}
where ${\cal L}_i=\sum_{|I|=i}CH_{I}$ with $CH_{I}=(\textbf{M}_{I}+
\hat{\textbf{M}}_{I})\textbf{M}_{\overline{I}}
+(\textbf{M}_{I}-\hat{\textbf{M}}_{I})
\hat{\textbf{M}}_{\overline{I}}$. Similar to Appendixes B and C, we can prove that Results S1 and S2 holds for any $n$-partite connected quantum network consisted of generalized EPR states and GHZ states.

For an odd $n$, from the proof in Appendix C, the quantum network of ${\cal N}_q$ can be locally transformed into two new networks ${\cal N}''_q$ and ${\cal N}'''_q$ satisfying that each party in ${\cal N}''_q$ or ${\cal N}'''_q$ has odd number of particles. For the network of ${\cal N}''_q$, the genuinely multipartite nonlocality can be verified by the inequality (\ref{eq5s}) with four inputs. Similar result holds for ${\cal N}'''_q$. Note that for each party $\textsf{B}_j$, there are four inputs as shown in Eq.(\ref{BB}) if local states of $\textsf{B}_j$ are same as each other in both networks ${\cal N}''_q$ and ${\cal N}'''_q$. Otherwise, there are at most eight inputs associated with two sets of observables as shown in Eq.(\ref{BB}). For some special network, it may be reduced to six inputs, where there are two different observables for verifying ${\cal N}'''_q$ and ${\cal N}'''_q$ (see the following example). Note that the number of inputs for each party depends on the transformed networks ${\cal N}''_q$ or ${\cal N}'''_q$, which is from the network configuration of ${\cal N}_q$. We cannot get the explicit form of the inequality (\ref{eq5s}) in this case.

\textbf{Continuing Example S1}. Consider the triangle network consisting of three EPR states by using the inequality (\ref{C1}). Assume that two parties of Alice, Bob and Charlie share one generalized EPR state: $|\phi_i\rangle_{2i-1,2i}=\cos\theta_i|00\rangle+\sin\theta_i|11\rangle$ with $\theta_i\in (0, \pi)$ and $i=1, 2$, and $3$.
The total system is denoted as $|\phi_1\rangle_{12}|\phi_2\rangle_{34}|\phi_3\rangle_{56}$.  Alice owns the particles 1 and 6. Bob owns the particles 2 and 3 while Charlie owns the particles 4 and 5. Here, $|{\phi}_1\rangle$ and $|{\phi}_3\rangle$ will be changed into four-particle GHZ states with local operations and auxiliary qubits. Define local observables as
\begin{eqnarray}
&&A_1=B_3=C_3=\cos\theta\sigma_z^{\otimes 3}+\sin\theta\sigma_x^{\otimes 3},
\nonumber\\
&&A_2=B_4=C_4=\cos\theta\sigma_z^{\otimes 3}-\sin\theta\sigma_x^{\otimes 3},
\nonumber\\
&&A_3=B_1=\sigma_z^{\otimes 3},
A_4=B_2=\sigma_x^{\otimes 3},
\nonumber\\
&&B_5=\sigma_z^{\otimes 4},
B_6=\sigma_x^{\otimes4},
\nonumber\\
&&C_1=\sigma_z\otimes \sigma_z,
C_2=\sigma_x\otimes \sigma_x,
\end{eqnarray}
From a forward evaluation we get
\begin{eqnarray}
\varpi_q&:=&\langle{}A_1B_1C_1\rangle+\langle{}A_2B_1C_1\rangle+\langle{}A_1B_2C_2\rangle
-\langle{}A_2B_2C_2\rangle
\nonumber\\
&&+\langle{}A_3B_3C_1\rangle+\langle{}A_3B_4C_1\rangle+\langle{}A_4B_3C_2\rangle
-\langle{}A_4B_4C_2\rangle
\nonumber\\
&&+\langle{}A_3B_5C_3\rangle+\langle{}A_3B_5C_4\rangle+\langle{}A_4B_6C_3\rangle
-\langle{}A_4B_6C_4\rangle
\nonumber\\
&=&6\cos\theta+6\sin\theta\sin2\theta_1 \sin2\theta_2\sin2\theta_3
\nonumber\\
&>&4\sqrt{2}+2
\label{G7}
\end{eqnarray}
which violates the inequality (\ref{C1}) when $\theta_i$s satisfy
\begin{eqnarray}
\sin2\theta_1 \sin2\theta_2\sin2\theta_3>0.7928
\end{eqnarray}

Moreover, consider the noisy network ${\cal N}_q$ given by
\begin{eqnarray}
\rho=\rho_{1}\otimes\rho_{2}\otimes\rho_{3}
\end{eqnarray}
where $\rho_{i}$ denotes Werner state \cite{Wrner} given by $\rho_{i}=v|\phi_i\rangle\langle\phi_i|+\frac{1-v}{4}\mathbbm{1}$ with $\theta_i=\frac{\pi}{4}$, and $v\in (0,1]$. The visibility is given by $v=0.966$. This is going beyond a recent result without the robustness \cite{23}.

\textbf{Continuing Example S2}. Assume that two parties of Alice, Bob and Charlie share a generalized EPR state: $|\phi_i\rangle_{2i-1,2i}=\cos\theta_i|00\rangle+\sin\theta_i|11\rangle$ with $\theta_i\in (0, \pi)$ and $i=1, 2$, and $3$.
The total system is denoted as $|\phi_1\rangle_{12}|\phi_2\rangle_{34}|\phi_3\rangle_{56}$. Alice owns the particles 1 and 6. Bob owns the particles 2 and 3 while Charlie owns the particles 4 and 5. To verify the genuinely tripartite nonlocality by using the inequality (\ref{C2}), Alice and Bob will change $|\phi_1\rangle_{12}$ into four-particle GHZ state $|\phi_1\rangle_{12ab}=\cos\theta_1|0000\rangle+\sin\theta_1|1111\rangle$ with local two-particle unitary operations for the inputs $3$ and $4$. Define observables of three observers as follows:
\begin{eqnarray}
&&A_1=\cos\theta\sigma_z\otimes \sigma_z+\sin\theta\sigma_x\otimes \sigma_z,
\nonumber\\
&&
A_2=\cos\theta\sigma_z\otimes \sigma_z-\sin\theta\sigma_x\otimes \sigma_z,
\nonumber\\
&&A_3=\sigma_z^{\otimes 3}, A_4=\sigma_x^{\otimes 3},
\nonumber\\
&&B_1=\sigma_z\otimes \sigma_z, B_2=\sigma_x\otimes \sigma_x,
\nonumber\\
&&B_3=\cos\theta\sigma_z^{\otimes 3}+\sin\theta\sigma_x^{\otimes 3},
\nonumber\\
&&
B_4=\cos\theta\sigma_z^{\otimes 3}-\sin\theta\sigma_x^{\otimes 3},
\nonumber\\
&&C_1=\sigma_z\otimes \sigma_z, C_2=\sigma_x\otimes \sigma_z,
\nonumber\\
&&C_3=\cos\theta\sigma_z\otimes \sigma_z+\sin\theta\sigma_x\otimes \sigma_z,
\nonumber\\
&&
C_4=\cos\theta\sigma_z\otimes \sigma_z-\sin\theta\sigma_x\otimes \sigma_z,
\label{G6}
\end{eqnarray}
From a forward evaluation we get
\begin{eqnarray}
\varpi_q&:=&\langle{}A_1B_1C_1\rangle+\langle{}A_2B_1C_1\rangle+\langle{}
A_1B_2C_2\rangle
\nonumber\\
&&-\langle{}A_2B_2C_2\rangle+\langle{}A_3B_3C_1\rangle
+\langle{}A_3B_4C_1\rangle
\nonumber\\
&&+\langle{}A_4B_3C_2\rangle-\langle{}A_4B_4C_2\rangle
+\langle{}A_3B_1C_3\rangle
\nonumber\\
&&+\langle{}A_3B_1C_4\rangle+\langle{}A_4B_2C_3\rangle
-\langle{}A_4B_2C_4\rangle
\nonumber\\
&=&6\cos\theta+2\sin\theta(\sin2\theta_1\sin2\theta_2
\nonumber\\
&&+\sin2\theta_2\sin2\theta_3+\sin2\theta_1\sin2\theta_3)
\nonumber\\
&>&4\sqrt{2}+2
\label{G7}
\end{eqnarray}
which violates the inequality (\ref{C2}), where $\theta_i$s satisfy
\begin{eqnarray}
\sin2\theta_1\sin2\theta_2+\sin2\theta_2\sin2\theta_3+\sin2\theta_1\sin2\theta_3>
2\times{}2^{1/4}\approx 2.3784
\end{eqnarray}
For special case of $\theta_1=\theta_2=\theta_3$, it follows that  $\theta>\frac{1}{2}\arcsin((\frac{32}{81})^{1/8})\approx 0.1748\pi$, where $\theta$ satisfies $\cos\theta=\frac{1}{\sqrt{1+(\sin2\theta)^4}}\approx 0.59$. If $\theta_1=\theta_2=\frac{\pi}{4}$, it follows that $\theta_3>0.121\pi$. The same visibility can be followed from the inequalities (\ref{C1}) and (\ref{C2}).

\section{Robustness of the inequality (2)}

The goal in this section is to prove the result for noisy networks. Consider an $n$-partite connected network ${\cal N}_q$ consisting of noisy states  $\rho_i, \varrho_j$, where $\rho_i$ and $\varrho_i$ are Werner states \cite{Wrner} given respectively by
\begin{eqnarray}
&&\rho_i=v_i|\phi_i\rangle\langle\phi_i|+\frac{1-v_i}{4}\mathbbm{1}_4
\nonumber\\
&&\varrho_i=w_i|\psi_i\rangle\langle\psi_i|+\frac{1-w_i}{2^{k_i}}
\mathbbm{1}_{2^{k_i}}
\label{D2}
\end{eqnarray}
where $|\phi_i\rangle$ is EPR state, $|\psi_j\rangle$ is GHZ state with $k_j$ particles, $\mathbbm{1}_j$ is the identity operator with the rank $j$, $v_i,w_j \in(0,1]$, $i=1, \cdots, m_1; j=1, \cdots, m_2$.

\textbf{Result S3}. The network ${\cal N}_q$ is genuinely $n$-partite nonlocal if $v_i, w_j$ satisfy the following inequality:
\begin{eqnarray}
\prod_{i=1}^{m_1}\prod_{j=1}^{m_2}v_iw_j>
1-\frac{2-\sqrt{2}}{2N}
\label{D3}
\end{eqnarray}
with $N=2^{n-1}-1$.

\begin{figure}
\begin{center}
\resizebox{240pt}{180pt}{\includegraphics{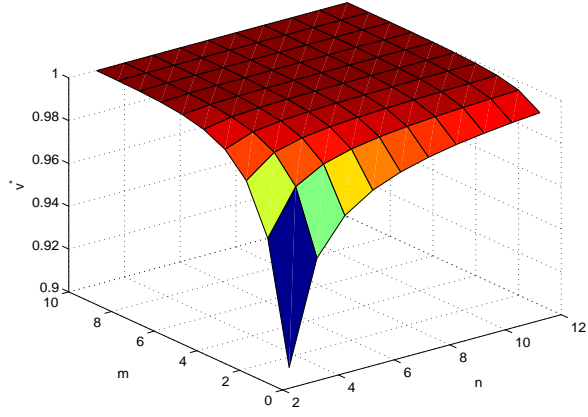}}
\end{center}
\caption{{\bf Visibility of Werner states}. Here, all $v_i$s and $w_j$s are equal to each other. $n$ denotes the number of parties while $m=m_1+m_2$ denotes the number of entangled states.}
\label{figS2}
\end{figure}

\textbf{Proof of Result S3}. The proof is similar to these stated in Appendixes B and C. Note that all observables are dichotomic. We get $CH_{I}'=\prod_{i=1}^{m_1} \prod_{j=1}^{m_2}v_iw_jCH_{I}$, where $CH_{I}$ is defined in the inequality (2) for the network ${\cal N}_q$ without noise while $CH'_{\vec{s}_i}$ denotes the same quantity associated with noisy network shown in Eq.(\ref{D2}). From Eqs.(\ref{B2})-(\ref{B6}) it implies that
\begin{eqnarray}
\varpi_q&=&\sum_{i}{\cal L}_i
\nonumber\\
&\geq &2N\sqrt{2}\prod_{i=1}^{m_1}\prod_{j=1}^{m_2}v_iw_j
\nonumber\\
&>&2^n\sqrt{2}-4\sqrt{2}+2
\label{D4}
\end{eqnarray}
when $v_i,w_j$ satisfy the inequality of $\prod_{i=1}^{m_1}\prod_{j=1}^{m_2}v_iw_j>1-\frac{2-\sqrt{2}}{2N}$. Numeric simulations are shown in Fig.S\ref{figS2} for the visibility of $n$-partite cyclic networks. $\Box$

\section{Proof of the inequality (3)}

In this section, we prove the inequality (3) for connected quantum networks. Consider a general network ${\cal N}$ (Fig.2) consisting of $n$ parties $\textsf{A}_1, \cdots, \textsf{A}_n$, who share $m$ sources $\lambda_1, \cdots, \lambda_m$. An eavesdropper may gain information related to the outcome of $\textsf{A}_i$ by measuring the correlated system with input $z_i$ and outcome $e_i$. Suppose that all the inputs of $\textbf{x}=x_1 \cdots{}x_n$ are accessible to the eavesdropper. Note that in secure applications all the legal parties will not send out their measurement outcomes before they begin to detect the eavesdropper. This means that a smart eavesdropper will recover one outcome $a_i$ by using the prior knowledge of $a_i,\textbf{x},z_i$, i.e., $P(e_i|a_i,\textbf{x},z_i)$. Hence, we consider the variation distance of two probability distributions $\{\prod_{i}P(e_i|a_i,\textbf{x},z_i)\}$ and $\{\prod_{i}P(e_i|\textbf{x},z_i)\}$. From the non-signalling condition we get that
\begin{eqnarray}
&&P(e_i|\textbf{x},z_i)=P(e_i|z_i),
        \nonumber
    \\
&&P(e_i|a_i,\textbf{x},z_i)=P(e_i|a_i,x_i,z_i)
\label{F1}%\label{eq25}
\end{eqnarray}
Denote $D_{e}:=D(\prod_{i=1}^nP(e_i|a_i; \textbf{x},z_i), \prod_{j=1}^n p(e_j|z_j))$ for convenience. From Eq.(\ref{F1}), the left side of the inequality (3) can be rewritten into

\begin{eqnarray}
D_e&\leq & \frac{1}{2}\sum_{\textbf{e}}p(e_1|a_1;
\textbf{x},z_1)|\prod_{i=2}^n p(e_i|a_i;\textbf{x},z_i)
-\prod_{i=2}^{n} p(e_i|z_i)|
\nonumber\\
&&
+\frac{1}{2}\sum_{\textbf{e}}|p(e_1|a_1;
\textbf{x},z_1)-p(e_1|z_1)|\prod_{i=2}^{n} p(e_i|z_i)
\label{F2}%\label{eq26}
\end{eqnarray}
Here, the inequality (\ref{F2}) is followed from the triangle inequality $|x-y|\leq |x-z|+|z-y|$. Moreover, we have

\begin{eqnarray}
D_e&\leq&D(p(e_1|a_1;{\textbf{x}}, z_1), p(e_1|z_1))
+D(\prod_{i=2}^np(e_i|a_i;
{\textbf{x}},z_i), \prod_{i=2}^np(e_i|z_i))
\label{F3}%\label{eq27}
\\
&\leq&\sum_{i=1}^nD(p(e_i|a_i;
{\textbf{x}}, z_i), p(e_i|z_i))
\label{F4}%\label{eq28}
\\
&=&\sum_{i=1}^nD(p(e_i|a_i; x_i,x_j, z_i), p(e_i|z_i))
\label{F5}%\label{eq30}
\\
&\leq & \sum_{i=1}^nI_2(P(a_i,a_j|x_i,x_j)
\label{F6}%\label{eq31}
\end{eqnarray}
Eq.(\ref{F3}) is from the normalization conditions of $\sum_{e_1}p(e_1|z_1)=1$, and $\sum_{e_j}p(e_j|a_j;{\textbf{x}}, z_j)=1$, $j=2, \cdots, n$. The inequality (\ref{F4}) is obtained by a similar process of the inequality (\ref{F2}) with $n-1$ iterations. Eq.(\ref{F5}) is followed from Eq.(\ref{F1}), where $\textsf{A}_i$ and $\textsf{A}_j$ shares at least one hidden variable. The inequality (\ref{F6}) is obtained by using the inequality of $D(p(e_i|a_i;x_i,x_j,z_i),p(e_i|z_i))\leq I_2(P(a_i,a_j|x_i,x_j))$ with the chained Bell inequality \cite{19,51}, where $I_2(P(a_i,a_j|x_i,x_j))=P(a_i=a_j|x_i=0;x_j=1)
+\sum_{|x_i-x_j|=1}P(a_i\not=a_j|x_i;x_j)$, $x_i$ and $a_i$ are the input and outcome of $\textsf{A}_i$, and $x_j$ and $a_j$ are the input and outcome of $\textsf{A}_j$.

Consider a quantum network ${\cal N}_q$ on which all parties have multiple inputs and binary outcomes. The quantum violation $\varpi_{es}$ of practical quantum correlations is shown as follows
\begin{eqnarray}
\varpi_{es}&=&\sum_{i=1}^{\lfloor\frac{n}{2}\rfloor}
\sum_{I}[\langle(\textbf{M}_{I}+
\hat{\textbf{M}}_{I})
\textbf{M}_{\overline{I}}\rangle
+\langle(\textbf{M}_{I}-
\hat{\textbf{M}}_{I})
\hat{\textbf{M}}_{\overline{I}}\rangle]
\nonumber\\
&\leq &\sum_{i=1}^{\lfloor\frac{n}{2}\rfloor}
    \sum_{I}[|\langle(\textbf{M}_{I}+
\hat{\textbf{M}}_{I})
\textbf{M}_{\overline{I}}\rangle|
\nonumber
+|\langle(\textbf{M}_{I}-
\hat{\textbf{M}}_{I})
\hat{\textbf{M}}_{\overline{I}}\rangle|]
\label{F7}%\label{eq33}
\\
&\leq &
\sum_{i=1}^{\lfloor\frac{n}{2}\rfloor}
\sum_{j\in I}[|\langle(M_{x_{j}(I)}
  +\hat{M}_{x_j(I)}){M}_{x_{k_j}(\overline{I})}\rangle|
+|\langle(M_{x_{j}(I)}
 -\hat{M}_{x_j(I)})\hat{{M}}_{x_{k_j}(\overline{I})}|]
\label{F8}%\label{eq33}
\\
&=&\sum_{i=1}^{\lfloor\frac{n}{2}\rfloor}
\sum_{I}CH_{M_{j}{M}_{k_j}}
\label{F9}%\label{eq34}
\end{eqnarray}
Here, the inequality (\ref{F7}) follows from the inequality of $\langle A\rangle\leq |\langle A\rangle|$. The inequality (\ref{F8}) is from the assumptions of separable observables ($M_{x_i}\leq 1$) except for $M_{x_{k_j}(\overline{I})}$ and $\hat{M}_{x_{k_j}(\overline{I})}$ given in Eq.(\ref{B52}). $M_{x_{k_j}}$ and $\hat{M}_{x_{k_j}}$ denote two observables of $\textsf{A}_{x_{k_j}}$ with $k_j\not\in {}I$ associated with $j\in I$. $CH_{M_{j}{M}_{k_j}}$ in Eq.(\ref{F9}) is defined by
\begin{eqnarray*}
CH_{M_{j}{M}_{k_j}}=
\langle(M_{x_{j}(I)}
  +\hat{M}_{x_j(I)}){M}_{x_{k_j}(\overline{I})}\rangle
+\langle(M_{x_{j}(I)}
 -\hat{M}_{x_j(I)})\hat{{M}}_{x_{k_j}(\overline{I})}\rangle
\end{eqnarray*}
where we have $\langle(M_{x_{j}(I)}
  +\hat{M}_{x_j(I)}){M}_{x_{k_j}(\overline{I})}\rangle\geq0$ and $\langle(M_{x_{j}(I)}
 -\hat{M}_{x_j(I)})\hat{{M}}_{x_{k_j}(\overline{I})}\rangle\geq0$ in applications (Appendixes B and C).

From the evaluations in Eqs.(\ref{B6}) and (\ref{B8}), we obtain that
\begin{eqnarray}
\varpi_{es}\leq \lfloor{}\frac{N}{n}\rfloor\sum_{j=1}^n
CH_{M_{j}{M}_{k_j}}
+2\sqrt{2}(\frac{N}{n}-\lfloor{}\frac{N}{n}\rfloor)
\label{F10}%\label{eq35}
\end{eqnarray}
where $N=2^n-1$. From $\langle XY\rangle=2p(X=Y)-1$ it follows that $I_2(P(a_i,a_j|x_i,x_j))=2-\frac{1}{2}CH_{X_{i}X_{j}}$. Using the inequalities (\ref{F6}) and (\ref{F10}), $D_e$ is evaluated as follows
\begin{eqnarray}
D_e\leq&2n-\frac{1}{2}
\sum_{j=1}^nCH_{M_jM_{k_j}}
\nonumber\\
\leq & 2n-\frac{\varpi_{es}-2\sqrt{2}
(n_0-\lfloor{}n_0\rfloor)}{2\lfloor{}n_0\rfloor}
\label{F11}%\label{eq36}
\end{eqnarray}
with $n_0=\frac{N}{n}$. This completes the proof.

\section{Classical simulation of quantum correlations with finite classical communication and shared randomness}

There are two assumptions for  guaranteeing the security against quantum or post-quantum eavesdroppers. One is that an eavesdropper may access all inputs, and the other is that an eavesdropper (not a party) cannot communicate with other parties. Otherwise, the bipartite quantum correlations can be precisely simulated with shared classical random variables \cite{34,36}. A similar result holds for multisource quantum networks. In fact, suppose that a general network ${\cal N}_q$ in Fig.2(b) is $k$-independent, that is, there are $k$ parties $\textsf{A}_1, \cdots, \textsf{A}_k$ (for simplicity) who do not share any entangled state with each other. ${\cal N}_q$ is schematically represented by a directed acyclic graph (DAG) (Fig.1(b)) with the same configuration as ${\cal N}_q$ using random variables $\lambda_1,  \cdots, \lambda_m$. $(\Omega_i,\Sigma_i,\mu_i)$ denotes the measure space of $\lambda_i$ with $i=1, \cdots, m$. Each observer $\textsf{A}_i$ obtains states from some random sources $\Lambda_{i}=\{\lambda_{j_{1}},  \cdots, \lambda_{j_{\ell_i}}\}$. The outcome $a_i$ of $\textsf{A}_i$ depends on the shared variables $\Lambda_i$ and the type of local measurement $x_i$. The joint conditional distributions of the outcomes conditional on the inputs are given
\begin{eqnarray}
P_c(\textbf{a}|\textbf{x})=\int_{\times_{i=1}^m\Omega_i} \prod_{j=1}^{n} P_c(a_j|x_j,\Lambda_{j})\prod_{i=1}^md\mu_i(\lambda_i)
\label{EE1}
\end{eqnarray}
Define a classical correlation as
\begin{eqnarray}
E_c(\vec{\textbf{x}})=\sum_{\Xi_1}
P_c(\textbf{a}_k|\vec{\textbf{x}}_k)-\sum_{\Xi_2}
P_c(\textbf{a}_k|\vec{\textbf{x}}_k)
\label{E2}
\end{eqnarray}
where $P_c(\textbf{a}_k|\vec{\textbf{x}}_k)$ denote the marginal probability of $\textbf{a}_k$ conditional on $\vec{\textbf{x}}_k$, and $\Xi_1$ denotes all combinations with even number of output $-1$ while $\Xi_2$ denotes all combinations with odd number of output $-1$, $\textbf{a}=a_1\cdots a_n$, $\textbf{x}=x_1\cdots x_n$, $\textbf{a}_k=a_1\cdots a_k$, and $\vec{\textbf{x}}_k=\vec{x}_1\cdots{}\vec{x}_k$.

Consider special applications with a joint system in the state $\rho$, such as multipartite QKD. All the independent parties of $\textsf{A}_1,  \cdots, \textsf{A}_{k}$ will finally perform qubit measurements after the others' local measurements and classical communication. The joint conditional probability $P(\textbf{a}|\vec{\textbf{x}}_k,x_{k+1}\cdots{}x_n)$ is defined
\begin{eqnarray}
P_q(\textbf{a}|\vec{\textbf{x}}_k,x_{k+1}\cdots{}x_n)={\rm tr}(\vec{x}_1\otimes \cdots \vec{x}_n \rho)
\end{eqnarray}
where $\vec{x}_i$ is a qubit measurement vector under the Pauli basis $(\sigma_x, \sigma_y, \sigma_z)$. Similar to Eq.(\ref{E2}), define a quantum correlation as
\begin{eqnarray}
E_q(\vec{\textbf{x}})=\sum_{\Xi_1}
P_q(\textbf{a}_k|\vec{\textbf{x}}_k)-\sum_{\Xi_2}
P_q(\textbf{a}_k|\vec{\textbf{x}}_k)
\label{EE2}
\end{eqnarray}

Consider a protocol completed in a device-independent manner as shown in Fig.3(b). Suppose that $\textsf{E}_i$s share variables $\Lambda_i$ consisting of independent random variables $\lambda_j$s with the same network configuration as ${\cal N}_q$, that is, all the parties $\textsf{A}_i$ are replaced with $\textsf{E}_i$ in the hidden variable model of ${\cal N}_q$. The goal in this section is to simulate the quantum correlation $E_q(\vec{\textbf{x}})$ in Eq.(\ref{EE2}) by using the classical correlation $E_c(\vec{\textbf{x}})$ in Eq.(\ref{E2}).

\textbf{Result S4}. Suppose that eavesdroppers $\textsf{E}_i$s can access all measurement inputs $\vec{\textit{x}}$. $\textsf{E}_i$s can generate a classical correlation $E_c(\vec{\textbf{x}})$ such that
\begin{eqnarray}
E_c(\vec{\textbf{x}})=E_q(\vec{\textbf{x}})
\label{E3}
\end{eqnarray}
by using shared randomness and finite classical communication, where the classical correlations are defined by $E_c(\vec{\textbf{x}})=\sum_{\Xi_1}
P_c(\textbf{e}_k|\vec{\textbf{x}}_k)-\sum_{\Xi_2}P_c(\textbf{e}_k|\vec{\textbf{x}}_k)$, $P_c(\textbf{e}_k|\vec{\textbf{x}}_k)$ denotes the joint distribution of eavesdroppers $\textsf{E}_1, \cdots, \textsf{E}_k$ conditional on the inputs $\vec{\textbf{x}}_k$, and their own inputs $\textbf{e}_k=e_1\cdots{}e_k$.

\textbf{Remark S2}. The classical simulation requires only finite classical communication for $\textsf{E}_i$s (see Fig.2(b)). Here, eavesdroppers cannot simulate the joint distribution $P({\bf a}|{\bf x})$ or the exact outputs ${\bf a}$ for each set of inputs ${\bf x}$. That is, the present simulation in Eq.(\ref{E3}) provides only potential information leakage of correlations based on generalized entanglement swapping going beyond chain-shaped networks without correlating assumptions \cite{36}. However, it does not imply the information leakage of measurement outputs in device-independent applications.

\begin{figure}
\begin{center}
\resizebox{480pt}{90pt}{\includegraphics{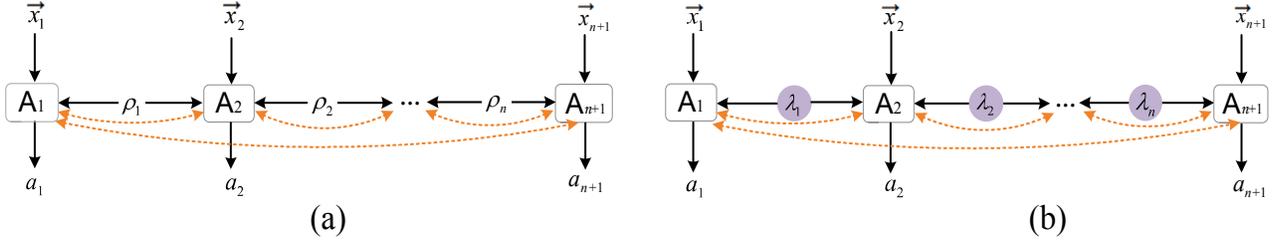}}
\end{center}
\caption{{\bf Schematically long-distance entanglement swapping network}. (a) Long-distance entanglement swapping network consisting of $k$ independent sources $\rho_i=|\phi_i\rangle\langle \phi_i|$ with generalized EPR states $|\phi_i\rangle=\alpha_i|01\rangle-\beta_i|10\rangle$. All the middle observers $\textsf{A}_2, \cdots, \textsf{A}_k$ can help $\textsf{A}_1$ and $\textsf{A}_{k+1}$ to build single entangled state with LOCC. (b) Classical simulation of correlations generated by a long-distance entanglement swapping experiment. Eavesdroppers share $n$ random variables $\lambda_1, \cdots, \lambda_k$. In simulation experiment, eavesdroppers can access to all measurement inputs and finite classical communication. The red dashed arrows represent to exchange classical messages.}
\label{figS3}
\end{figure}

\textbf{Proof of Result S4}.  Consider a general quantum network ${\cal N}_q$ (Fig.2(a) shown in the main text), which consists of all entangled pure states. Suppose that there are $k$ independent observers $\textsf{A}_1, \cdots, \textsf{A}_k$, who do not share entangled states. Our goal here is to show classical simulation of the multipartite correlations by eavesdroppers who can access all inputs and commutate with each other. A classical simulation with the same network configuration is shown in Fig.2(b) in the main text. One example of Fig.2 is the long-distance entanglement swapping network given in Fig.S\ref{figS3}(a). In experiment, all the middle observers $\textsf{A}_2, \cdots, \textsf{A}_k$ perform Bell measurements $\sigma_z\otimes \sigma_z, \sigma_x\otimes \sigma_x$ and send the outcomes to $\textsf{A}_1$ and $\textsf{A}_{k+1}$ who can recover a single entangled state with local unitary operations. To complete the simulation shown in Fig.S\ref{figS3}(b), suppose that $\textsf{A}_1$ and $\textsf{A}_{k+1}$ perform the qubit projective measurements represented by respective Bloch vector $\vec{x}_1, \vec{x}_{k+1}$ in terms of the Pauli basis $(\sigma_x,\sigma_y, \sigma_z)$. Suppose that $\textsf{A}_1$ and $\textsf{A}_{k+1}$ obtain binary outcomes $a_1, a_{k+1}=\pm 1$ respectively. The correlations of $\textsf{A}_1$ and $\textsf{A}_{k+1}$ exhibit the following form:
\begin{eqnarray}
E_q(\vec{x}_1,\vec{x}_{k+1})
:=&\sum_{\Xi_1}  P_q(\textbf{a}|\vec{\textbf{x}})
-\sum_{\Xi_2}P_q(\textbf{a}|\vec{\textbf{x}})
\label{E4}
\end{eqnarray}
where $P_q(\textbf{a}|\vec{\textbf{x}})$ is defined in Eq.(\ref{EE2}). An open problem \cite{3} is to determine the classical simulation capability of $E_q(\vec{x}_1, \vec{x}_{k+1})$ for $n\geq 3$ with independent sources and finite communication.

Generally, the proof of Eq.(\ref{E3}) is divided into three cases. One is to prove classical simulation of long-distance entanglement swapping shown in Fig.S\ref{figS3}(a) with EPR states for $k\geq 1$. The second is to prove classical simulation of generalized entanglement swapping for any acyclic networks consisting of EPR states. The last one is to prove the result for any acyclic networks consisting of generalized EPR states.

{\bf Case 1. Long-distance entanglement swapping on chain-shaped networks}

For $k=1$, the network shown in Fig.S\ref{figS3}(a) reduces to standard Bell network \cite{34}. For $k=2$, it is a standard entanglement swapping network that can be simulated classically \cite{36}. In what follows, we prove the result for $k>2$. The proof is completed by induction $k$.

\begin{figure}
\begin{center}
\resizebox{420pt}{95pt}{\includegraphics{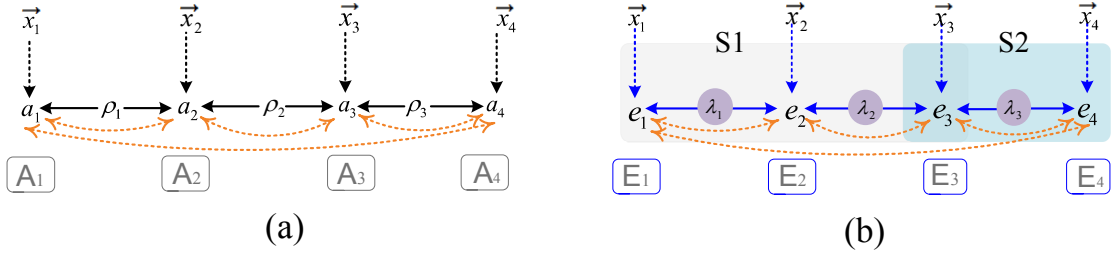}}
\end{center}
\caption{{\bf Classical simulation of quantum correlations from entanglement swapping}.  (a) Entanglement swapping with $4$ observers $\textsf{A}_1, \cdots, \textsf{A}_4$, who share EPR states $\rho_1\otimes \rho_2\otimes \rho_3$. (b) The classical simulation of quantum correlations generated by entanglement swapping by eavesdroppers $\textsf{E}_1, \cdots, \textsf{E}_4$ who share random variables $\lambda_1, \lambda_2, \lambda_3$. The classical communication is represented by red dashed arrows. The classical simulation is completed by two simulations S1 and S2 without assumptions of correlating sources.}
\label{figS4}
\end{figure}

We firstly prove the result for $k=3$, as shown in Fig.S\ref{figS4}. $\textsf{A}_i$ and $\textsf{A}_{i+1}$ share one or two EPR states shown in Fig.S\ref{figS4}(a). $\textsf{A}_1$ and $\textsf{A}_4$ obtain the respective measurement input $\vec{x}_1=(\cos\theta_a, 0, \sin\theta_a)$ and $\vec{x}_4=(\cos\theta_b, 0, \sin\theta_b)$. The correlation given in Eq.(\ref{E4}) is rewritten into
\begin{eqnarray}
{E}_q(\vec{x}_1,\vec{x}_4)
&=&\sum_{\Xi_1}
    P_q(\textbf{a}|\vec{\textbf{x}})
    -\sum_{\Xi_2}P_q(\textbf{a}|\vec{\textbf{x}})
    \nonumber
    \\
&=&\cos(\theta_a-\theta_b)
\label{E5}
\end{eqnarray}

The classical simulation is shown in Fig.S\ref{figS4}(b), where $\textsf{E}_i$ and $\textsf{E}_{i+1}$ share one or two random variables $\lambda_i$s with the same network configuration as shown in Fig.S\ref{figS4}(a). The finite classical communication is allowed for $\textsf{E}_i$s who know the inputs of $\textsf{A}_1$ and $\textsf{A}_4$. The simulation protocol consists of two separable simulations S1 and S2 as follows:
\begin{itemize}
\item[S1] $\textsf{E}_i$s set up the first classical simulation according to finite classical communication.
\begin{itemize}
\item[$\bullet$] $\textsf{E}_1, \textsf{E}_2$, and $\textsf{E}_3$ obtain a classical correlation as follows
    \begin{eqnarray}
   {E}_c(\vec{x}'_1, \vec{y}_1)&=&\sum_{\Xi_1}
    P_c(\textbf{e}|\vec{x}'_1, \vec{y}_1) -\sum_{\Xi_2}P_c(\textbf{e}|\vec{x}'_1, \vec{y}_1)
        \nonumber
    \\
&=&2\cos\theta_a
    \label{E6}
    \end{eqnarray}
    from a tripartite classical simulation protocol \cite{36} with independent sources and finite communication. Here, the inputs are assumed to be $\vec{x}'_1=(2\cos\theta_a, 0, 0)$ and $\vec{y}_1=(1, 0, 0)$.
\item[$\bullet$] $\textsf{E}_3$ and $\textsf{E}_4$ obtain a classical correlation as follows
    \begin{eqnarray}
    {E}_c(\vec{y}_1, \vec{x}'_4)&=&P_c(e_3= e_{4}|\vec{y}_1, \vec{x}'_4)
    -P_c(e_3\not=e_{4}|\vec{y}_1, \vec{x}'_4)
        \nonumber
    \\
&=&\cos\theta_b
    \label{E7}
    \end{eqnarray}
    from a bipartite classical simulation protocol \cite{34}, where  $\vec{x}'_4=(\cos\theta_b, 0, 0)$. Note that two steps are independent. It follows that
 \begin{eqnarray}
   {E}_c(\vec{x}'_1, \vec{x}'_4)&=&E_c(\vec{x}'_1,  \vec{y}_1)E_c(\vec{y}_1, \vec{x}'_4)
        \nonumber
    \\
&=&2\cos\theta_a\cos\theta_b,
    \label{E8}
 \end{eqnarray}
which can be easily followed from the separable Bell measurement of $\textsf{A}_3$, i.e., $\langle M_{x_1}M_{x_2}M_{x_3}M_{x_4}\rangle
=\langle{}M_{x_1}M_{x_2}M_{x_3;1}\rangle \langle{}M_{x_3;2}M_{x_4}\rangle$ with separable measurement $M_{x_3;1}\otimes M_{x_3;2}$ of $\textsf{A}_3$, where $M_{x_1}$ is observable with dichotomic outcomes. Equivalently, one can prove the result from the conditional independence of full correlations $P_q(\textbf{e}|\vec{x})=P_q(\textbf{e}_1|\vec{x}_1)P_q(\textbf{e}_2|\vec{x}_2)$, where $P_q(\textbf{e}_1|\vec{x}_1)$ and $P_q(\textbf{e}_2|\vec{x}_2)$ denote the respective correlation obtained from the tripartite network (consisting of $\textsf{E}_1, \textsf{E}_2$, and $\textsf{E}_3$) and the bipartite network (consisting of $\textsf{E}_3$ and $\textsf{E}_4$). Here, $\textsf{E}_3$ in Fig.S\ref{figS4}(b) or $\textsf{A}_3$ in Fig.S\ref{figS4}(a) performs the separable measurement.
\end{itemize}
\item[S2] $\textsf{E}_i$s set up the second classical simulation according to finite classical communication.
\begin{itemize}
\item[$\bullet$] $\textsf{E}_1, \textsf{E}_2$, and $\textsf{E}_3$ obtain a classical correlation as follows
    \begin{eqnarray}
   {E}_c(\vec{x}''_1,\vec{y}_1)&=&\sum_{\Xi_1}
    P_c(\textbf{e}|\vec{x}''_1, \vec{y}_1)
    -\sum_{\Xi_2}P_c(\textbf{e}|\vec{x}''_1, \vec{y}_1)
        \nonumber
    \\
&=&2\sin\theta_a
    \label{E9}
    \end{eqnarray}
    from a tripartite classical simulation protocol \cite{36} by using independent sources and finite communication, where their inputs are assumed to be $\vec{x}''_1=(0,0, 2\sin\theta_a)$ and $ \vec{y}_1=(0, 0, 1)$.
\item[$\bullet$] $\textsf{E}_3$ and $\textsf{E}_4$ obtain a classical correlation as follows
    \begin{eqnarray}
   {E}_c(\vec{y}_1, \vec{x}''_4)&=&P_c(e_3= e_{4}|\vec{y}_1, \vec{x}''_4)
    -P_c(e_3\not=e_{4}|\vec{y}_1, \vec{x}''_4)
    \nonumber
    \\
&=&\sin\theta_b
    \label{E10}
    \end{eqnarray}
from a bipartite classical simulation protocol \cite{34} without correlating sources, where $\vec{x}''_4=(0,0, \sin\theta_b)$. These two steps are independent. It follows that
 \begin{eqnarray}
   {E}_c(\vec{x}''_1, \vec{x}''_4)&=&
   {E}_c(\vec{x}''_1, \vec{y}_1)
   {E}_c(\vec{y}_1, \vec{x}''_4)
    \nonumber
    \\
    &=&2\sin\theta_a\sin\theta_b
    \label{E11}
 \end{eqnarray}
\end{itemize}
\end{itemize}

From Eqs.(\ref{E8}) and (\ref{E11}), it follows that
\begin{eqnarray}
\sum_{\Xi_1}
    P_c(\textbf{e}|
    \vec{\textbf{x}})
    -\sum_{\Xi_2}P_c(\textbf{e}
|\vec{\textbf{x}})
&:=&\frac{1}{2}({E}_c(\vec{x}'_1,\vec{x}'_4)
+{E}_c(\vec{x}''_1, \vec{x}''_4))
\nonumber\\
&=&\cos(\theta_a-\theta_b)
\nonumber\\
&=&{E}_q(\vec{x}_1, \vec{x}_4)
\label{E12}
\end{eqnarray}
from Eq.(\ref{E5}), where S1 and S2 are two independent simulations which are recombined into one simulation with equal probability. No correlating operation is required for any sources. In this protocol, total classical communication is finite because each simulation uses finite communication.

Now, by induction we can prove the result for any $k\geq 3$, where the classical simulations are divided into two separable chained subnetworks consisting of $\textsf{E}_1, \cdots, \textsf{E}_k$, and $\textsf{E}_k$ and $\textsf{E}_{k+1}$, respectively. Similar result holds for generalized measurement in terms of the Pauli basis \cite{34,36} when all measurement inputs are accessible to eavesdroppers without assumption of correlating sources.

\begin{figure}
\begin{center}
\resizebox{440pt}{160pt}{\includegraphics{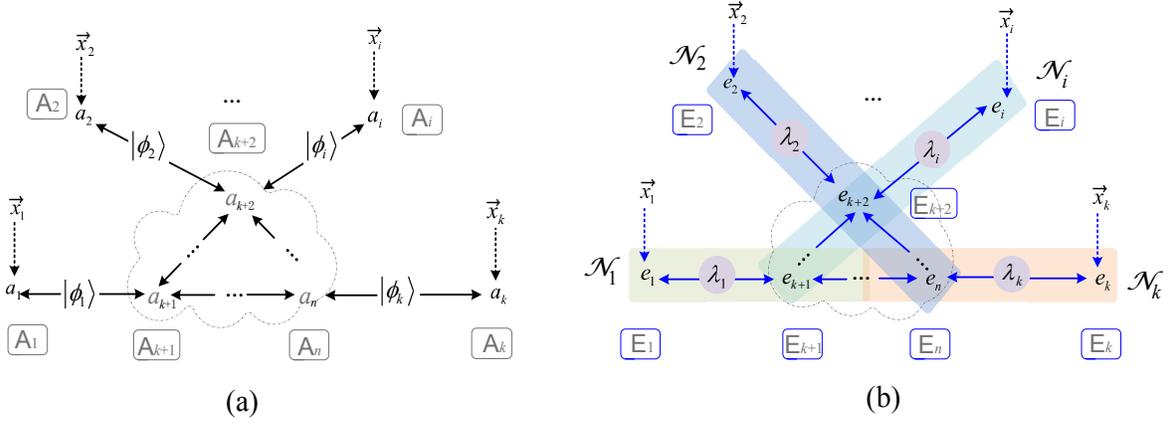}}
\end{center}
\caption{{\bf Classical simulation of quantum correlations from generalized entanglement swapping}. (a) A generalized entanglement swapping on any acyclic connected quantum network ${\cal N}_q$. The resources consist of generalized EPR states $\otimes_i|\phi_i\rangle$ with $|\phi_i\rangle=\alpha_i|01\rangle+\beta_i|10\rangle$. $\textsf{A}_{k+1}, \cdots, \textsf{A}_n$ perform the multiple-particle Bell measurement on local systems and send outcomes to other observers $\textsf{A}_1, \cdots, \textsf{A}_{k}$, who can recover a multipartite generalized GHZ state by proper local unitary operations. (b) The classical simulation of generalized entanglement swapping by non-signalling eavesdroppers $\textsf{E}_1,\cdots, \textsf{E}_n$ who share variables $\lambda_i$s. The red dashed arrows represent exchanging classical messages. The classical simulation network is decomposed into $k$ chain-shaped subnetworks ${\cal N}_1, \cdots, {\cal N}_k$. Finite classical communication is allowed for all eavesdroppers.}
\label{figS5}
\end{figure}

{\bf Case 2. Generalized entanglement swapping on any connected acyclic quantum networks}

Consider a connected acyclic quantum network ${\cal N}_q$ consisting of EPR states $|\phi_i\rangle=\frac{1}{\sqrt{2}}(|00\rangle+|11\rangle)$, as shown in Fig.S\ref{figS5}(a). One goal of ${\cal N}_q$ is that all parties $\textsf{A}_1, \cdots, \textsf{A}_k$ who have no prior-shared entanglement create a multipartite GHZ state with the help of other parties $\textsf{A}_{k+1}, \cdots, \textsf{A}_n$. In experiment, each party $\textsf{A}_{k+j}$ performs a multi-particle Bell measurement on the local system and sends outcome for other parties $\textsf{A}_i$s, who can recover a $k$-particle GHZ state $|GHZ\rangle=\frac{1}{\sqrt{2}}(|0\rangle^{\otimes k}+|1\rangle^{\otimes k})$ by performing proper local unitary operations. Suppose that all parties $\textsf{A}_i$s finally obtain binary outcomes $a_i\in \{-1, 1\}$ after performing local qubit measurements. Here, $\textsf{A}_i$s perform the single qubit measurement with measurement input $\vec{x}_i=(\cos\theta_i, 0, \sin\theta_i)$ on Bloch sphere. Their outcomes have correlations as follows:
\begin{eqnarray}
E_q(\vec{\textbf{x}}) &=&\sum_{\Xi_1}P_q(\textbf{a}|\vec{\textbf{x}})
-\sum_{\Xi_2}P_q(\textbf{a}|\vec{\textbf{x}})
        \nonumber
    \\
&=&\prod_{i=1}^k\cos\theta_i
+\prod_{i=1}^k\sin\theta_i
\label{E13}
\end{eqnarray}
where $\vec{\textbf{x}}=\vec{x}_1\cdots\vec{x}_k$ are measurement vectors under the Pauli basis of single qubit system.

In what follows, we show that these correlations can be simulated classically, as shown in Fig.S\ref{figS5}(b) with even $k$. Similar proof holds for odd $k$. The classical simulation is completed by powerful eavesdroppers $\textsf{E}_i$s who can access all the measurement inputs with fully independent variables $\lambda_j$s. The proof is completed by the following separable steps:
\begin{itemize}
\item[S1] Eavesdroppers divide the simulation network shown in Fig.S\ref{figS5}(b) into $n$ chain-shaped subnetworks ${\cal N}_1, \cdots, {\cal N}_k$ with the help of classical communication, where each subnetwork ${\cal N}_i$ contains
    $\textsf{E}_i$ and $\textsf{E}_{k+j}$, and all subnetworks are connected and ${\cal N}_i\cap{\cal N}_j\not=\varnothing$.
\item[S2] For the subnetwork ${\cal N}_1$, eavesdroppers set up the first simulation according to finite classical communication. They obtain classical correlation as follows
    \begin{eqnarray}
    {E}_c^1&=&\sum_{\Xi_1}
    P_c(\textbf{e}_1|\vec{\textbf{x}}_1)
    -\sum_{\Xi_2}P_c(\textbf{e}_1
    |\vec{\textbf{x}}_1)
        \nonumber
    \\
&=&2\cos\theta_{1}
    \label{E14}
    \end{eqnarray}
from the multipartite classical simulation protocol given in Case 1, where $\vec{\textbf{x}}_1$ and $\textbf{e}_1$ denote the respective inputs and outcomes of all eavesdroppers included in the subnetwork ${\cal N}_1$, the input of the eavesdropper $\textsf{E}_1$ is given by $\vec{x}'_1=(2\cos\theta_1, 0, 0)$ while the inputs of other eavesdroppers are given by $(1, 0, 0)$. Note that this simulation procedure does not require to correlate variables $\lambda_i$s.
\item[S3] By repeating S2 for all subnetworks ${\cal N}_i$, eavesdroppers can obtain classical correlation as follows
    \begin{eqnarray}
    {E}_c^i&=&\sum_{\Xi_1}
    P_c(\textbf{e}_i|\vec{\textbf{x}}_i)
-\sum_{\Xi_2}P_c(\textbf{e}_i|\vec{\textbf{x}}_i)
        \nonumber
    \\
&=&\cos\theta_{i}
    \label{E15}
    \end{eqnarray}
    for $i=2,\cdots, n$, where $\vec{\textbf{x}}_i$ and $\textbf{e}_i$ denote the respective inputs and outcomes of eavesdroppers included in the subnetwork ${\cal N}_i$, the input of the eavesdropper $\textsf{E}_i$ is given by $\vec{x}'_i=(\cos\theta_i, 0, 0)$ while the inputs of all the other eavesdroppers are given by $(1, 0, 0)$ in the subnetwork ${\cal N}_i$. Note that these steps are independent. It implies that
 \begin{eqnarray}
    E_c(\vec{\textbf{x}}')
    &=&\sum_{\Xi_1}P_c(\textbf{e}|\vec{\textbf{x}}')
-\sum_{\Xi_2}P_c(\textbf{e}|\vec{\textbf{x}}')
          \nonumber
    \\
&=&\prod_{i=1}^n{E}_c^i
          \nonumber
    \\
&=&2\prod_{i=1}^k\cos\theta_i
    \label{E16}
 \end{eqnarray}
   which can be followed from the Bell observables of $\textsf{A}_{k+j}$s shown in Fig.S\ref{figS5}(b), where $\vec{\textbf{x}}'$ and $\textbf{e}$ denote the respective inputs and outputs of eavesdroppers included in ${\cal N}_q$.
\item[S4] Eavesdroppers set up the second classical simulation with the help of finite classical communication. They obtain multipartite correlations
    \begin{eqnarray}
   \hat{E}_c^i&=&\sum_{\Xi_1}
    P_c(\textbf{e}'_i|\vec{\textbf{x}}''_i)
-\sum_{\Xi_2}P_c(\textbf{e}'_i|\vec{\textbf{x}}''_i)
        \nonumber
    \\
&=&\sin\theta_i
    \label{E17}
    \end{eqnarray}
for $i\geq 2$ and $2\sin\theta_1$ for $i=1$ from the classical simulation protocol given in Case 1, where the input of $\textsf{E}_i$ is given by $\vec{x}''_i=(0,0,\sin\theta_i)$ for $i\geq 2$ or $\vec{x}''_1=(0,0,2\sin\theta_1)$ while the inputs of other eavesdroppers are given by $(0, 0, 1)$ in the subnetwork ${\cal N}_i$, $i=1, \cdots, n$. Note that these steps are independent. It follows that
  \begin{eqnarray}
    \hat{E}_c(\vec{\textbf{x}}'')&=&\prod_{i=1}^k\hat{E}_c^i
            \nonumber
    \\
&=&  2\prod_{i=1}^k\sin\theta_i
    \label{E18}
 \end{eqnarray}

Finally, from Eqs.(\ref{E16}) and (\ref{E18}) it implies that
\begin{eqnarray}
\sum_{\Xi_1}
    P_c(\textbf{e}|\vec{\textbf{x}})
    -\sum_{\Xi_2}P_c(\textbf{e}|\vec{\textbf{x}})
&:=&\frac{1}{2}(\hat{E}_c(\vec{\textbf{x}}')
+\hat{E}_c(\vec{\textbf{x}}''))
        \nonumber
    \\
&=&\prod_{i=1}^k\cos\theta_i
+\prod_{i=1}^k\sin\theta_i
        \nonumber
    \\
&=&E_q(\vec{\textbf{x}})
\label{E19}
\end{eqnarray}
\end{itemize}
from Eq.(\ref{E13}), where two independent simulations are recombined into one simulation with equal probability. Similar result holds for generalized measurement in terms of the Pauli basis \cite{34,36}.

\begin{figure}[ht]
\begin{center}
\resizebox{340pt}{190pt}{\includegraphics{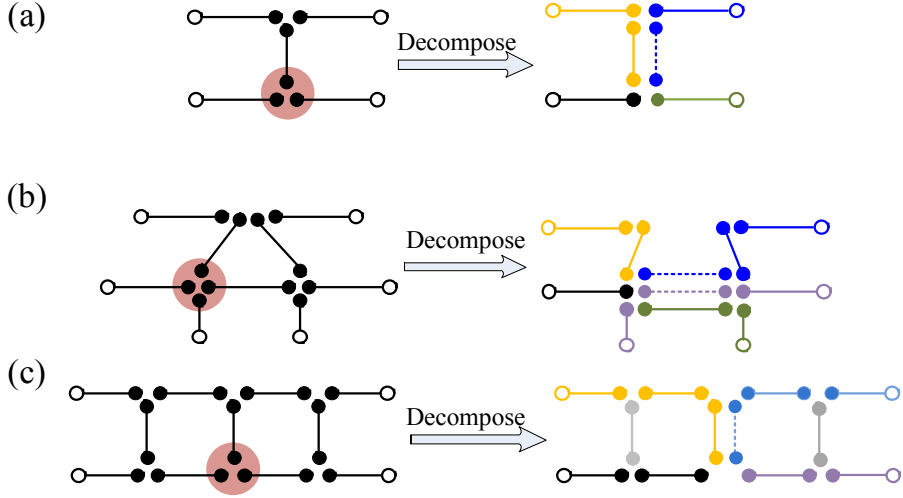}}
\end{center}
\caption{{\bf Decomposing simulation networks}. (a) An acyclic network consisting of five random variables. (b) A network with one cycle consisting of 9 random variables. (c) A network with two cycles consisting of 11 random variables. Each variable is represented by two dots (or one dot and one circle) connected by one line. All the empty circles denote independent observers who want to create an entanglement in terms of the quantum network. The shadowed node denotes the root of the equivalent tree. The right side of each subnetwork consists of several decomposed simulation networks with several chain-shaped subnetworks, where each subnetwork is represented by one colored path. The dotted lines denote the reused random variables. The grey figures are useless for generalized entanglement swapping experiment. The proof holds for special cyclic networks shown in Fig.S5(b) and Fig.S5(c), where all the independent agents are not included in any cycle.}
\label{figS6}
\end{figure}

{\bf Case 3. Entanglement swapping on acyclic networks consisting of generalized EPR states}

Consider an $n$-partite connected acyclic quantum network ${\cal N}_q$ consisting of generalized EPR states $|\psi_i\rangle=\alpha_i|00\rangle+\sqrt{1-\alpha_i^2}|11\rangle$, as shown in Fig.S\ref{figS5}(a). One goal of ${\cal N}_q$ is to help all parties $\textsf{A}_i$s to build a generalized multipartite GHZ state assisted by local measurements of other parties $\textsf{A}_{k+j}$, $i=1, \cdots, k$; $j=1, \cdots, n-k$. In experiment, each party $\textsf{A}_{k+j}$ performs a multi-particle Bell measurement on the local systems and sends outcome to $\textsf{A}_i$s who can recover a generalized $k$-partite GHZ state $|GHZ\rangle=\gamma|0\rangle^{\otimes k}+\sqrt{1-\gamma^2}|1\rangle^{\otimes k}$ with success probability $p$ by performing proper local unitary operations, where $p$ and $\gamma$ depend on $\alpha_i$s and $\beta_i$s. Note that the probability distribution $p$ can be generated by one party in a classical simulation experiment. So, it is sufficient to prove the result for one generalized GHZ state $|GHZ\rangle$. Suppose that $\textsf{A}_i$s obtain binary outcomes $a_i\in\{-1, 1\}$. In the classical simulation, $\textsf{A}_i$s finally perform the single qubit measurement with measurement input $\vec{x}_i=(\cos\theta_i, 0, \sin\theta_i)$. Their outcomes for even $k$ exhibit nonlocal correlations of the following form:
\begin{eqnarray}
E_q(\vec{\textbf{x}})&=&\sum_{\Xi_1}P_q(\textbf{a}
|\vec{\textbf{x}})
-\sum_{\Xi_2}P_q(\textbf{a}|\vec{\textbf{x}})
\nonumber
\\
&=&2\gamma\sqrt{1-\gamma^2}\prod_{i=1}^k\cos\theta_i
+\prod_{i=1}^k\sin\theta_i
\label{E20}
\end{eqnarray}

Consider a classical network ${\cal N}$ consisting of $m$ variables $\lambda_i$s with the similar network configuration as ${\cal N}_q$. Suppose that ${\cal N}$ is shared by classical eavesdroppers $\textsf{E}_i$s. For classical simulation, eavesdroppers can perform a probabilistic simulation with an input $(2\gamma\sqrt{1-\gamma^2}, 0, 0)$. From a similar procedure of Case 2, it is easy to prove that $E(\vec{\textbf{x}})$ given in Eq.(\ref{E20}) can be simulated classically using several independent simulations by eavesdroppers who can access to all measurement inputs.

Before closing the proof, we show that S1 in Case 2 can be completed with finite classical communication, where other steps require finite classical communication from Case 2. Note that ${\cal N}$ is acyclic \cite{Graph}. There is one node that is connected to each agent $\textsf{E}_i$ with only one path. This path is contained in the subnetwork ${\cal N}_i$. In theory, an acyclic graph is equivalent to a tree where the root is the desired center node, and some edges are allowed to be in different paths, i.e., one random variable can be used more than one time in classical simulation. Some examples are shown in  Fig.S\ref{figS6}(a). Generally, a similar result holds for special cyclic networks shown in  Fig.S\ref{figS6}(b) and Fig.S\ref{figS6}(c), where all the independent agents are not included in any cycle. This completes the proof. $\Box$

\section{Proof of the inequality (4)}

From the inequality (2), the expect violation of quantum correlations is represented by
\begin{eqnarray}
\varpi_{es}&=&\sum_{a_i,b_i,c_i\atop{x_i,y_i,z_i\in {S}_{in}}}[(-1)^{a_1+b_1+c_1+x_1y_1z_1}
P(a_1b_1c_1|x_1y_1z_1)
\nonumber
\\
&&+(-1)^{a_2+c_2+b_2+x_2y_2z_2}
P(a_2b_2c_2|x_2y_2z_2)
+(-1)^{a_3+b_3+c_3+x_3y_3z_3}
P(a_3b_3c_3|x_3y_3z_3)]
\label{G1}%\label{eq37}
\end{eqnarray}
where $P(a_ib_ic_i|x_iy_iz_i)=\textrm{tr}
[(A^{a_i}_{x_i}B^{b_i}_{y_i}C^{c_i}_{z_i})\rho]$, $A^{a_i}_{x_i}$, $B^{b_i}_{y_i}$ and $C^{c_i}_{z_i}$ denote the POVM operators of $\textsf{A}_1$, $\textsf{A}_2$ and $\textsf{A}_3$, respectively, $\rho$ denotes the total state of the triangle network (Fig.3), and ${S}_{in}$ denotes the input set satisfying the conditions $x_1=y_1$, $x_2=z_2$ and $y_3=z_3$. From the symmetry of inputs $x_i$, $y_i$, and $z_i$, it follows that
\begin{eqnarray}
\varpi_{es}=3\!\!\!\!\!\!\sum_{a_1,b_1,c_1\atop{x_1,y_1,z_1\in {S}_{in}}}\!\!\!\!\!\!(-1)^{a_1+b_1+c_1+x_1y_1z_1}
P(a_1b_1c_1|x_1y_1z_1)
\label{G2}
%\label{eq38}
\end{eqnarray}

In what follows, we only consider the case of $a_1=0$ and $x_1=0$ to prove the following inequality
\begin{eqnarray}
P(a_1|x_1)\leq \frac{1}{2}+\frac{1}{12}\sqrt{72-\varpi^2_{es}}
\label{G3}%\label{eq39}
\end{eqnarray}
which is inspired by bipartite QKD scheme \cite{20,52}. Note that the right side of the inequality (4) is concave function of $\varpi_{es}$. Denote $W=\sum_{a_1,b_1,c_1\atop{x_1,y_1,z_1 \in {\cal {}S}_{in}}}(-1)^{a_1+b_1+c_1+x_1y_1z_1}A^{a_1}_{x_1}B^{b_1}_{y_1}C^{c_1}_{z_1}$ as tripartite quantum operator. Inspired by the semidefinite programming relaxation of the bipartite scheme \cite{20,52}, from Eq.(\ref{G2}) the inequality (\ref{G3}) is equivalent to the following operator inequality:
\begin{eqnarray}
A^{0}_0\leq \frac{1}{2}
+\frac{6}{\sqrt{72-\varpi^2}}
-\frac{\varpi}{12\sqrt{72-\varpi^2}}W
\label{G4}%\label{eq40}
\end{eqnarray}
This operator inequality is then rewritten into
\begin{eqnarray}
\frac{1}{2}+\frac{1}{2}A_0\leq \frac{1}{2}+\frac{6}{
\sqrt{72-\varpi^2}}
-\frac{\varpi}{12\sqrt{72-\varpi^2}}W
\label{G5}%\label{eq41}
\end{eqnarray}
where $A_0=A^{0}_0-A^{1}_0$ is Hermitian operator.

Now, it is sufficient to prove the operator inequality (\ref{G5}) for the guessing probability in Eq.(\ref{G3}). Define four operators $L_i$s as follows \cite{20,51}:
\begin{eqnarray}
L_1&=&-\frac{\sqrt{\gamma}}{2}A_1
+\frac{\varpi\sqrt{\gamma}}{24}(B_0C_0-B_1C_1)
+\frac{\varpi}{48\sqrt{\gamma}}(A_1B_0C_0+A_1B_1C_1),
\\
L_2&=&\frac{\sqrt{\gamma}}{8}
-\frac{\varpi}{24\sqrt{\gamma}}A_0
+\frac{1}{8\sqrt{\gamma}}(B_0C_0+B_1C_1)
\nonumber\\
&&-\frac{\sqrt{\gamma}}{2}(A_0B_0C_0+A_0B_1C_1)
+\frac{(36-\varpi^2)\sqrt{\gamma}}{144}(A_1B_0C_0-A_1B_1C_1)
\\
L_3&=&\frac{\sqrt{\gamma}}{4}(A_0B_0C_0-A_0B_1C_1)
-\frac{1}{8\sqrt{\gamma}}(B_0C_0-B_1C_1)
\nonumber\\
&&-\frac{(36-\varpi^2)\sqrt{\gamma}}{144}
(A_1B_0C_0+A_1B_1C_1)
\end{eqnarray}

\begin{eqnarray}
L_4&=&\frac{1}{4\sqrt{\gamma}}
-\frac{(36-\varpi^2)\sqrt{\gamma}}{72}A_0-\frac{\sqrt{\gamma}\varpi}{24}(B_0C_0+B_1C_1)
\nonumber\\
&&
-\frac{\varpi}{48\sqrt{\gamma}}(A_1B_0C_0-A_1B_1C_1)
\end{eqnarray}
where $\gamma=\frac{3}{\sqrt{72-\varpi^2}}$. It is from a forward evaluation to check that
\begin{eqnarray}
\sum_{i=1}^4L^\dag_iL_i&=&\frac{6}{\sqrt{72-\varpi^2}}
-\frac{1}{2}A_0
+\frac{\varpi}{12\sqrt{72-\varpi^2}}
W
\label{G6}
%\label{eq42}
\\
&\geq &0
\label{G7}
%\label{eq43}
\end{eqnarray}
for $\varpi<6\sqrt{2}$. Eq.(\ref{G6}) makes use of the equalities $A_x^2=B_y^2=C_z^2=\mathbbm{1}$, and $[A_x,B_y]=[A_x,C_z]=[B_y,C_z]=0$ (commute  relations). The inequality (\ref{G7}) follows from the fact that the right side of Eq.(\ref{G6}) is positive semidefinite, i.e., $\sum_{i=1}^4L^\dag_i L_i\geq0$. Finally, the inequality (\ref{G5}) is obtained from the concavity of its right side for $\varpi$ with $\varpi\in [4\sqrt{2}+2,6\sqrt{2}]$.

\section{Proof of the inequality (5)}

The length of the real secret $k$ is lower bounded by $H_{min}(A|E)-s_{p}$ (up to terms of order $\sqrt{s}$), where $H_{min}(A|E)$ denotes the min-entropy \cite{20,39} of $A$ depending on Eve's information, $s_{p}\approx sH(A|B)$ is used for error correcting \cite{20,39}, $H(A|B)=-\sum_{a,b}p(a,b)\log_2p(a|b)$ denotes the conditional entropy depending on the average probability with specific outcomes $a,b$ being observed. Hence, it is essentially to get the asymptotic secret-key rate $R$ in Eq.(5) by evaluating $H_{min}(A|E)$ \cite{20}. After the local measurement of $\textsf{A}_1$, the joint state of $\textsf{A}_1$ and Eve can be represented by $\rho_{AE}=\sum_{\bf a}P({\bf a}|{\bf x}_{r})|{\bf a}\rangle\langle {\bf a}|\otimes \rho_{E}^{\bf a}$, where $\rho_{E}^{\bf a}$ denotes the reduced state owned by Eve conditional on $\textsf{A}_1$'s output ${\bf a}$. Let $H_{min}({\bf a}|E)$ be the min-entropy \cite{20,39} of ${\bf a}$ depending on Eve's information related to $\rho_{AE}$.

Now, we estimate the guessing probability of eavesdropper Eve about the raw key ${\bf a}=a_1\cdots{}a_s$ using $s$ pairs of the triangular network (Fig.3). Let $P({\bf a}|{\bf x}_{r},ez)$ denote the probability of ${\bf a}$ conditional on Eve's information. The expected optimal guessing probability for Eve is shown as follows
\begin{eqnarray}
P_{gs}({\bf a}|E)=\max_z\sum_{e}P(e|z)\max_{\bf a}P({\bf a}|{\bf x}_{r},ez)
\label{H1}%\label{eq44}
\end{eqnarray}

Denote $\rho_{ABC|ez}$ as the initial state when Eve measures $z$ and obtains the outcome $e$, where $\rho_{ABC|ez}$ satisfies $\rho_{ABC}=\sum_eP(e|z)\rho_{ABC|ez}$. Denote ${\bf A}^{\bf a}_{{\bf x}_{r}}=\prod_{i=1}^sA^{a_i}_{x_{r}}$ as POVMs for all outputs ${\bf a}$ conditional on $\textsf{A}_1$'s inputs ${\bf x}_{r}$. We get that
\begin{eqnarray}
P(\textbf{a}|\textbf{x}_{r},ez)=\textrm{tr}
[\rho_{ABC|ez}^{\otimes{}s}{\bf A}^{\textbf{a}}_{\textbf{x}_{r}}]
\label{H2}
%\label{eq45}
\end{eqnarray}
where ${\bf x}_{r}$ denote all parties' inputs with $x_r$ for each experiment. By using the semidefinite programming relaxation \cite{20,52}, it follows that
\begin{eqnarray}
{\bf A}^{\textbf{a}}_{\textbf{x}_{r}}\leq \textbf{A}(\varpi)=\prod_{i=1}^s[\alpha(\varpi)
\mathbbm{1}+\beta(\varpi)O_i]
\label{H3}
%\label{eq46}
\end{eqnarray}
where $\alpha(\varpi)=\frac{1}{2}+\frac{1}{12}\sqrt{72-\varpi^2}
+\frac{\varpi^2}{12\sqrt{72-\varpi^2}}$, $\beta(\varpi)=-\frac{\varpi}{12\sqrt{72-\varpi^2}}$ and $O_i=\sum_{abc,xyz}g^{abc}_{xyz}A^{a_i}_{x_i}
B^{b_i}_{y_i}C^{c_i}_{z_i}$. From the inequality (\ref{H3}) we obtain that
\begin{eqnarray}
P_{gs}(\textbf{a}|E)&=&\max_z\sum_{e}P(e|z)\max_{\textbf{a}}
\rm{tr}[\rho_{ABC|ez}^{\otimes{}s}{\bf A}^{\textbf{a}}_{\textbf{x}_{r}}]
\nonumber\\
&\leq & \max_z\sum_{e}P(e|z)
 \min_{\varpi}
 \textrm{tr}[\rho_{ABC|ez}^{\otimes{}s}\textbf{A}(\varpi)]
\nonumber\\
&\leq & \min_{\varpi}\textrm{tr}[\rho_{ABC}^{\otimes{}s}\textbf{A}(\varpi)]
\nonumber
\\
&\leq&\min_{\varpi}
 (\alpha(\varpi)+\beta(\varpi)\varpi_{es}+s^{-1/4}_{es})^s
\label{H4}
%\label{eq47}
\\
&\leq &(\frac{1}{2}+\frac{1}{12} \sqrt{72-\varpi_{es}^2}+s^{-1/4}_{es})^s
\label{H5}
%\label{eq48}
\end{eqnarray}
The inequality (\ref{H4}) follows from the result \cite{20,52} by combining the systems B and C, where $s_{es}$ denotes the number of the binary key distilled from the raw key. The inequality (\ref{H5}) follows from the inequality
\begin{eqnarray}
\min_\varpi\{\alpha(\varpi)+\beta(\varpi)\varpi_{es}\}
\leq \frac{1}{2}+\frac{1}{12}\sqrt{72-\varpi_{es}^2}
\end{eqnarray}
using the concavity.

Denote $R:=\frac{H_{min}({\bf a}|E)}{s}=-\frac{\log_2P_{gs}({\bf a}|E)}{s}$ as the asymptotic secret-key rate of the tripartite QKD scheme \cite{20,39}. It follows from the inequality (\ref{H5}) that
\begin{eqnarray}
R&=&-\frac{\log_2P_{gs}({\bf a}|E)}{s}
\nonumber\\
&\geq& -\log_2(\frac{1}{2}+\frac{1}{12}
\sqrt{72-\varpi_{es}^2})-H(a|b,c)
\label{H6}
%\label{eq48s}
\end{eqnarray}
where $P_{gs}({\bf a}|E)$ denotes the expected optimal guessing probability conditional on the eavesdropper's input $E$.

\section{Verifying general quantum networks with LOCC}

\begin{figure}[ht]
\begin{center}
\resizebox{260pt}{80pt}{\includegraphics{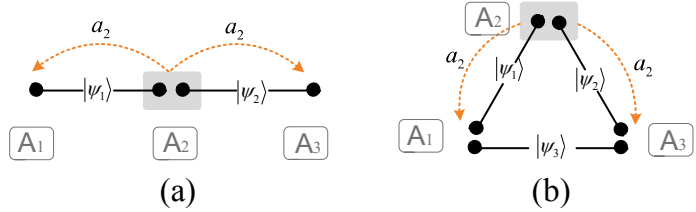}}
\end{center}
\caption{{\bf Entanglement swapping network}. (a) An entanglement swapping network consisting of three parties. (b) A triangle network consisting of three parties. $|\psi_i\rangle$ are bipartite entangled pure states. }
\label{figS7}
\end{figure}

In this section, we verify general quantum networks with LOCC. Take the network shown in Fig.S\ref{figS7} as an example. For Fig.S\ref{figS7}(a), each pair of $\textsf{A}_1$ and $\textsf{A}_{2}$, or $\textsf{A}_2$ and $\textsf{A}_{3}$ shares one bipartite entangled pure state $|\psi_i\rangle$. Note that any two parties, $\textsf{A}_2$ and $\textsf{A}_{3}$ for example, can recover a new bipartite entangled state $|\phi_{2}\rangle$ when the other party $\textsf{A}_1$ performs a two-particle joint measurement and sends out outcomes. Now, $\textsf{A}_2$ and $\textsf{A}_{3}$ share one generalized bipartite entangled state $|\phi_i\rangle$. It means that any two parties can rebuild a bipartite entangled state assisted by LOCC. Moreover, consider the network as shown in Fig.S\ref{figS7}(b), where each pair of $\textsf{A}_1$ and $\textsf{A}_{2}$, $\textsf{A}_2$ and $\textsf{A}_{3}$, and $\textsf{A}_3$ and $\textsf{A}_{1}$ shares one bipartite entangled pure state $|\psi_i\rangle$. Any two adjacent parties ($\textsf{A}_2$ and $\textsf{A}_{3}$ for example) can recover a new bipartite entangled state $|\psi_{4}\rangle$ when the other ($\textsf{A}_1$ in example) performs a two-particle joint measurement and sends out outcomes. $\textsf{A}_2$ and $\textsf{A}_{3}$ share two generalized bipartite entangled states $|\psi_2\rangle |\psi_4\rangle$. So, any two parties rebuild a bipartite entangled state assisted by LOCC \cite{28}. This fact is useful for detecting the genuinely multipartite nonlocality of all the multipartite entangled pure states using tailed CHSH inequalities. Interestingly, it is also applicable for network scenarios. Here, the detecting experiment consists of two steps. One is to construct one bipartite entangled state for each pair with LOCC. The other is to verify a generalized bipartite network.

\textbf{Result S5}. The genuinely multipartite nonlocality of any connected quantum networks consisting of genuinely entangled multipartite pure states can be verified with the help of LOCC.

\textbf{Proof of Result S5}. The proof is divided into three steps inspired by a recent result for single entangled system \cite{28}. The first is to prove that any genuinely multipartite entangled pure states can be useful for entanglement swapping. The second is to prove any two observers in a connected network consisting of genuinely multipartite entangled pure states can recover a new bipartite entangled pure state with the help of others' LOCC. The last is to verify that any two parties with shared $k$ bipartite entangled pure states using the maximal violation of some bipartite Bell inequality.

{\bf Lemma S2}. Consider a $4$-partite network consisting of Alice, Bob, Charlie and Tom, where Alice, Bob and Tom share one genuinely entangled pure state $|\psi_1\rangle$, and Bob and Charlie and Tom share the other genuinely entangled pure state $|\psi_2\rangle$. Alice and Charlie can create a new bipartite entangled pure state with the help of LOCC of Bob and Tom.

Compared with the standard tripartite entanglement swapping with two EPR states \cite{8s}, there are other parties (Tom, one party or multiple parties) which can be viewed as controllers. The proof is completed by two steps. One is to obtain two bipartite entangled pure states for Alice and Bob (Bob and Charlie) with the help of LOCC of Tom. The other is to use tripartite entanglement swapping \cite{8s}. Similarly, we have

{\bf Lemma S3}. Consider a $k$-partite connected network consisting of $\textsf{A}_1, \cdots, \textsf{A}_k$, where $\textsf{A}_i$ shares at least one genuinely multipartite entangled pure state $|\psi_i\rangle$ with other parties. Any two parties who do not share any entangled state can create a new bipartite entangled pure state with the help of LOCC of other parties.

Now, we continue to prove the result. Consider an $n$-partite connected network ${\cal N}$ consisting of $\otimes_i|\psi_i\rangle$, where $|\psi_i\rangle$ is genuinely multipartite entangled pure state on Hilbert space $\otimes_{j=1}^{s_i}\mathcal{H}_j$. The proof is similar to its for single-source network \cite{7s}. For any pair of parties $\textsf{A}_{i_1}$ and  $\textsf{A}_{i_2}$, two cases will be discussed. One is that $\textsf{A}_{i_1}$ and  $\textsf{A}_{i_2}$ share one entangled pure state \cite{9s}. The other is that $\textsf{A}_{i_1}$ and  $\textsf{A}_{i_2}$ do not share any entangled pure state.

{\bf Step 1}. ${\cal N}_q$ has $n$-partite genuinely nonlocality if there exists a covering set such that each pair in it can create entanglement in all branches of LOCC protocol for other parties.

By using the tilted-CHSH family of Bell inequalities \cite{6s,7s,10s}, there is a linear Bell inequality defined by
\begin{eqnarray}
B_{i_1i_2}=\sum_{{\bf s}}\sum_{\bf a, x}B^{({\bf s})}_{{\bf x}; {\bf a}}\leq C
\label{I1}
\end{eqnarray}
when at least one pair of $\textsf{A}_{i_1}$ and  $\textsf{A}_{i_2}$ cannot create one bipartite entangled pure state. Here, $B_{i_1i_2}$ is used to verify the final entangled pure state for $\textsf{A}_{i_1}$ and $\textsf{A}_{i_2}$, and $B^{({\bf s})}_{{\bf x}; {\bf a}}$ depends on the specific branch ${\bf s}$ and inputs ${\bf a}$ and outputs ${\bf x}$ of all involved parties \cite{6s,7s,10s}. Otherwise, the achievable quantum upper bound is $C'>C$. Here, we choose $C, C'$ to be independent of $i_1i_2$ \cite{6s,7s}. Define
\begin{eqnarray}
{\cal B}=\sum_{(i_1,i_2)}B_{i_1i_2}
\label{I2}
\end{eqnarray}
where the summation is taken over all pairs in a covering set.

Assume that the total state of ${\cal N}_q$ is a biseparable state in Eq.(\ref{bs}). From the definition of LOCC, the joint system shared by  $\textsf{A}_{i_1}$ and $\textsf{A}_{i_2}$ which are in different partitions remain in a separable state in each branch of the protocol \cite{7s}. So the maximum of $B_{i_1i_2}$ is at most $C$ if $i_1\in I$ and $i_2\in \overline{I}$. It implies that $\max{\cal B}=(k-1)C'+C<kC'$ for any biseparable state $\rho_{bs}$ while $\max{\cal B}=kC'$ for a fully entangled ${\cal N}_q$.

{\bf Step 2}. ${\cal N}_q$ consisting of generalized EPR states and genuinely entangled GHZ states satisfies the assumption in Step 1.

Consider an $n$-partite connected network ${\cal N}_q$ consisting of generalized EPR states $\otimes_{i=1}^{m_1}|\phi_i\rangle$ and genuinely entangled GHZ states $\otimes_{j=1}^{m_2}|\psi_j\rangle$. Since ${\cal N}_q$ is connected, $|\Omega\rangle:=\otimes_{i=1}^{m_1}\otimes_{j=1}^{m_2}
|\phi_i\rangle|\psi_j\rangle$ can be viewed as an $n$-partite entangled pure state with local dimension $d$, where $d$ is the maximal dimension of all
local systems. Otherwise, there exist at least two parties who cannot share an entangled pure state with LOCC. From Lemmas 1 and 2, it means that two parties are not connected, i.e., the network is disconnected. A simple fact of these networks is as follows: for each $\textsf{A}_i$, there are projections on local systems such that the remained network of the other $n-1$ parties also consists of EPR states and GHZ states with LOCC. Here, the measurement outcomes are sent to $n-2$ parties. After these operations, any pair of parties can share at least one generalized EPR state with LOCC of other $n-2$ parties.

{\bf Step 3}. Bell test for two parties with shared bipartite entangled pure states $\otimes_{i}|\phi_i\rangle$.

For a connected network, any pair of parties $\textsf{A}_i, \textsf{A}_j$ can create $k_{ij}$ bipartite entangled pure states $\otimes_{s=1}^{k_{ij}}|\phi_s\rangle$, where $|\phi_s\rangle=\cos\theta_s|00\rangle+\sin\theta_s|11\rangle$. We can rewrite the total state into
\begin{eqnarray}
\otimes_{s=1}^{k_{ij}}|\phi_s\rangle=
\sum_{t=0}^{2^k-1}\alpha_t|tt\rangle
\label{I4}
\end{eqnarray}
where $\sum_{t=0}^{2^{k_{ij}}-1}\alpha_t|tt\rangle=:|G\rangle$ is a bipartite state on Hilbert space $\mathbb{C}^{2^{k_{ij}}}\times \mathbb{C}^{2^{k_{ij}}}$, and $\alpha_t$'s are coefficients depending on $\theta_s$'s. There is a tailored Bell inequality which will be maximally violated by this state. One example is the tailed CHSH \cite{6s} for verifying the nonlocality of $|G\rangle$ for some $\theta_i\not=0$. Another method is using a linear Bell inequality with $3$ and 4 inputs and $2^{k_{ij}}$ outputs \cite{10s}.

{\bf Step 4}. ${\cal N}_q$ consisting of all genuinely entangled pure states satisfies the assumption in Step 1.

For each pair of $\textsf{A}_i$ and $\textsf{A}_j$, there exists a generalized chain-shaped subnetwork consisting of $\textsf{A}_i, \textsf{A}_{s_1}, \cdots, \textsf{A}_{s_\ell}, \textsf{A}_j$. For each entangled pure state $\rho_{1\cdots{} m}$, there exist local measurements on the $m$-th particle such that the resultant is entangled pure state $\varrho_{1\cdots{}m-1}$ \cite{7s}. By using these local operations for all the other parties (iteratively for each multipartite state), they can recover a standard chain-shaped subnetwork consisting of all $d$-dimensional bipartite entangled pure states $|\Phi_j\rangle$. Similar to Lemma S3, $\textsf{A}_i$ and $\textsf{A}_j$ can recover at least one bipartite entangled pure state using the local $d$-dimensional rotations after $\textsf{A}_{s_1}, \cdots, \textsf{A}_{s_\ell}$ perform local joint measurement under the generalized Bell basis: $\{|B_{st}\rangle=\frac{1}{\sqrt{d}}
\sum_{j=0}^{d-1}\exp(\frac{2isj\pi}{d})|j\rangle
|d-j\rangle\}$. Finally, the nonlocality of the resultant can be verified by $\textsf{A}_i$ and $\textsf{A}_j$ using the Bell inequality with multiple setting \cite{10s}. $\square$

\textbf{Result S6}. Consider a connected $n$-partite network ${\cal N}_q$ consisting of EPR and GHZ states with white noise. ${\cal N}_q$ has genuinely $n$-partite nonlocality with the help of LOCC if each noisy state can be verified with a Bell inequality.

\textbf{Proof}. Firstly, consider the entanglement swapping network consisting of Alice, Bob and Charlie, who share two Werner states \cite{Wrner}:
\begin{eqnarray}
\rho_i=v_i|\phi\rangle\langle \phi|+\frac{1-v_i}{4}\mathbbm{1}_4, i=1,2
\label{I5}
\end{eqnarray}
where $|\phi\rangle$ is EPR state, $\mathbbm{1}_4$ is the identity operator with the rank $4$, and $v_i\in (0,1]$. In what follows, we consider the resultant of Alice and Charlie with the help of LOCC of Bob. Specially, Bob performs a two-qubit joint measurement under Bell basis: $\{\frac{1}{\sqrt{2}}(|00\rangle\pm|11\rangle), \frac{1}{\sqrt{2}}(|01\rangle\pm|10\rangle)\}$, and sends the outcome to Alice or Charlie, who can then recover a bipartite state as follows:
\begin{eqnarray}
\rho=&v_1v_2|\phi\rangle\langle\phi|
+\frac{v_1+v_2-2v_1v_2}{2}
(|00\rangle\langle 00|+|11\rangle\langle 11|)
+\frac{(1-v_1)(1-v_2)}{4}\mathbbm{1}_4
\label{I6}
\end{eqnarray}
The fidelity of $\rho$ with respect to EPR state $|\phi\rangle$ is given by
\begin{eqnarray}
F(\rho,|\phi\rangle\langle\phi|)
=&{\rm tr}[|\phi\rangle\langle\phi|\rho]=\frac{v_1+v_2}{2}
\label{I7}
\end{eqnarray}
which satisfies $\min\{v_1,v_2\}\leq \frac{v_1+v_2}{2}\leq \max\{v_1,v_2\}$. It means that $\rho$ can be verified when both $\rho_1$ and $\rho_2$ can be verified using proper Bell test.

Moreover, consider a $k$-particle GHZ state with white noise as follows \cite{Wrner}:
\begin{eqnarray}
\rho_{1\cdots{} k}=v|\psi\rangle\langle \psi|+\frac{1-v}{2^k}\mathbbm{1}_{2^k}
\label{I8}
\end{eqnarray}
which is shared by $k$ parties, where $|\psi\rangle=\frac{1}{\sqrt{2}}(|0\rangle^{\otimes k}+|1\rangle^{\otimes k})$, and $\mathbbm{1}_{2^k}$ is the identity operator with rank $2^k$, and $v\in (0,1]$. One party performs local measurement under the basis $\{\frac{1}{\sqrt{2}}(|0\rangle\pm|1\rangle)\}$ and sends the outcome to the others who can recover a $k-1$-particle GHZ state with white noise, i.e., $\rho_{1\cdots{} k-1}=v|\psi'\rangle\langle \psi'|+\frac{1-v}{2^{k-1}}\mathbbm{1}_{2^{k-1}}$, where $|\psi'\rangle$ is GHZ state with $k-1$ particles given by $|\psi'\rangle=\frac{1}{\sqrt{2}}
(|0\rangle^{\otimes k-1}+|1\rangle^{\otimes k-1})$, and $\mathbbm{1}_{2^{k-1}}$ is the identity operator with the rank $2^{k-1}$.

With two facts stated above, we can prove the result. The only difference for quantum networks consisting of pure states is that the final bipartite states are noisy EPR states $\varrho_1, \cdots, \varrho_k$. Assume that $\rho_1, \cdots, \rho_m$ are input noisy states, where $v_1, \cdots, v_m$ are entanglement fractions. Since each noisy state $\rho_i$ can be verified using some Bell inequality ${\cal L}_i\leq 0$, $\otimes_j\varrho_j$ can be verified similarly using the Bell inequality ${\cal B}_t\leq 0$, where $t$ satisfies $v_t=\min\{v_1,\cdots, v_m\}$ from the first fact given above. $\square$

\textbf{Example S4}. A useful application of these results is to verify quantum resources for measure-based quantum computation, as shown in Fig.4. Especially, consider universal resources of hypergraph states $|E\rangle=\prod_{\{i_1, \cdots, i_k\}\in{}E}CZ_{i_1i_2\cdots{}i_k}|+\rangle^{\otimes n}$, where $\{i_1, \cdots,i_k\}$ denote $k$ vertices connected by a $k$-hyperedge in edge set $E$ \cite{39}, and $CZ_{i_1\cdots{}i_k}$ are controlled-phase gates among the $k$ connected qubits. One example is W-type state $\frac{1}{2}(|000\rangle+|010\rangle+|100\rangle+|111\rangle)$ from $3$-hyperedge that is inequivalent to GHZ state from a chain-shaped graph under LOCC. The genuinely multiparticle nonlocality of single hypergraph state is verified using Hardy-type arguments or Bell inequalities \cite{12s}. Result S5 proves the genuinely multipartite nonlocality of connected networks consisting of graph states and hypergraph states. It provides a general witness of hybrid resources for measurement-based quantum computation.

\section{Device-independent secret sharing with triangle network}

The main idea is inspired by \cite{MBNC} with GHZ state and Svetlichny inequality \cite{4}. Our consideration is to use triangle network with the present inequality (2). Consider the triangle network shown in Fig.3 shared by Alice, Bob and Charlie. Here, the inequality (\ref{C1}) given by
\begin{eqnarray}
\varpi&=&A_1B_1C_1+A_2B_1C_1+A_1B_2C_2-A_2B_2C_2
\nonumber\\
&&+A_3B_3C_1+A_3B_4C_1+A_4B_3C_2-A_4B_4C_2
\nonumber\\
&&+A_3B_5C_3+A_3B_5C_4+A_4B_6C_3-A_4B_6C_4
\nonumber\\
&\leq & 4\sqrt{2}+2
\nonumber\\
\label{J1}
\end{eqnarray}
will be used to construct a (3,2) DIQSS. Here, assume that Alice is untrusted and try to recover the output of other parties. Note that the maximal violation of this inequality is given by $\Delta=6\sqrt{2}$, that is, $A_1B_1C_1+A_2B_1C_1+A_1B_2C_2-A_2B_2C_2=2\sqrt{2}$, $A_3B_3C_1+A_3B_4C_1+A_4B_3C_2-A_4B_4C_2=2\sqrt{2}$ and $A_3B_5C_3+A_3B_5C_4+A_4B_6C_3-A_4B_6C_4=2\sqrt{2}$. This implies that Bob and Charlie is maximally entangled for quantum settings or shares a PR-box in the non-signaling settings even if they know the input and output of Alice. Hence, Bob and Charlie can ensure that their reduced states are uncorrelated with any other system if they can get the maximal violation  of the inequality (\ref{J1}). In this case, the guess probability of the output $a_2$ or $a_3$ is given by $p(e=a_2|x_1,x_2,x_3)=p(e=a_3|x_1,x_2,x_3)=1/2$, that is, randomly guess the output. This means that the maximal violation rules out the guess attack of an untrusted part. In applications, the secure scheme can be obtained with large violation. The main idea is solving linear optimization given in Eqs.(6)-(11) by using the NPA hiearchy \cite{51,MBNC}. Eve's guess probability $p(e=c|x_1,x_2,x_3)$ in quantum settings is given by $p(e=c|x_1,x_2,x_3)\leq 3/4$ for $\varpi\geq 8.25$ and $P_{guess}=1/2$ for $\varpi=6\sqrt{2}$. Note that the inequality (\ref{J1}) is not symmetric. This implies a larger violation $\varpi\geq 8.31$ required for $p(e=b|x_1,x_2,x_3)\leq 3/4$ for guessing the output $b$. These are large than recent scheme \cite{MBNC} using GHZ state and Svetlichny inequality with less inputs \cite{4s}. Generally, the perfect security (in terms of random guess) can be ensured by the maximal violation of the present inequality. This provides a useful method for featuring the quantum secret sharing.


\begin{thebibliography}{99}
\bibitem{1}  J. S. Bell, On the Einstein-Podolsky-Rosen paradox. \textit{Phys}. \textbf{1}, 195 (1964).

\bibitem{2}  A. Einstein,  B. Podolsky, and N.  Rosen, Can quantum-mechanical description of physical reality be considered complete? \textit{Phys. Rev.} \textbf{47}, 777-780 (1935).

\bibitem{3}  L. Masanes, All bipartite entangled states are useful for information processing. \textit{Phys. Rev. Lett.} \textbf{96}, 150501 (2006).

\bibitem{4} N. Brunner,  D. Cavalcanti,  S. Pironio,   V. Scarani, and   S. Wehner, Bell nonlocality. \textit{Rev. Mod. Phys}. \textbf{86}, 419 (2014).

\bibitem{5}A. K. Ekert,  Quantum cryptography based on Bell's theorem. \textit{Phys. Rev. Lett.} \textbf{67}, 661-663 (1991).

\bibitem{6}  D. Mayers and  A.  Yao, Quantum cryptography with imperfect apparatus. \textit{Proc. 39th Annual Symp. Found. Comput. Sci. (FOCS)} 503-512 (1998).

\bibitem{7} J. Barrett, L. Hardy, and A. Kent, No signaling and quantum key distribution.  \textit{Phys. Rev. Lett}. \textbf{95}, 010503 (2005).

\bibitem{8}  U. Vazirani and   T. Vidick, Fully device independent quantum key distribution. \textit{Phys. Rev. Lett}. \textbf{113}, 140501 (2014).

\bibitem{9} S. Pironio, A. Ac\'{\i}n, S. Massar, A. Boyer de la Giroday, D. N. Matsukevich, P. Maunz, S. Olmschenk, D. Hayes, L. Luo, T. A. Manning \& C. Monroe, Random numbers certified by Bell's theorem. \textit{Nature} \textbf{464}, 1021 (2010).

\bibitem{10}R. Colbeck and R. Renner, Free randomness can be amplified. \textit{Nat. Phys}. \textbf{8}, 450 (2012).

\bibitem{11} R. Gallego, L. Masanes, G. D. L. Torre, C.  Dhara, L.  Aolita, and  A. Ac\'{i}n, Full randomness from arbitrarily deterministic events. \textit{Nat. Commun}. \textbf{4}, 2654 (2013).

\bibitem{12}L.-M. Duan,  M. Lukin,  J. I. Cirac and  P. Zoller, Long-distance quantum communication with atomic ensembles and linear optics. \textit{Nature} \textbf{414}, 413 (2001).

\bibitem{13} H. J. Kimble, The quantum Internet. \textit{Nature} \textbf{453}, 1023 (2008).

\bibitem{14}  C. M. Lee and  R. W. Spekkens, Causal inference via algebraic geometry: feasibility tests for functional causal structures with two binary observed variables. \textit{J. Causal Inference} \textbf{5}, 20160013 (2017).

\bibitem{15} C. Branciard, N. Gisin and   S. Pironio, Characterizing the nonlocal correlations created via entanglement swapping. \textit{Phys. Rev. Lett}. \textbf{104}, 170401 (2010).

\bibitem{16a}T. Fritz, Beyond Bell's theorem: correlation scenarios. \textit{New J. Phys.} \textbf{14}, 103001 (2012).

\bibitem{16b}A. Tavakoli, P. Skrzypczyk, D. Cavalcanti, and A. Ac\'{\i}n, Nonlocal correlations in the star-network configuration. \textit{Phys. Rev. A} \textbf{90}, 062109 (2014).

\bibitem{16} D. Rosset, C. Branciard, T. J. Barnea, G. P\"{u}tz, N. Brunner, and N. Gisin, Nonlinear Bell inequalities tailored for quantum networks. \textit{Phys. Rev. Lett}. \textbf{116}, 010403 (2016).


\bibitem{17} R. Chaves, Polynomial bell inequalities. \textit{Phys. Rev. Lett}. \textbf{116}, 010402 (2016).

\bibitem{18}M.-X. Luo, Computationally efficient nonlinear Bell inequalities for quantum networks. \textit{ Phys. Rev. Lett.} \textbf{120}, 140402 (2018).

\bibitem{luo}M. X. Luo, A nonlocal game for witnessing quantum networks. \textit{npj Quant. Inf.} \textbf{5}, 91 (2019).

\bibitem{18b}N. Gisin, J.-D. Bancal, Y. Cai, P. Remy, A. Tavakoli, E. Zambrini Cruzeiro, S. Popescu, and N. Brunner, Constraints on nonlocality in networks from no-signaling and independence, \textit{Nat. Commun.} \textbf{11}, 2378 (2020).

\bibitem{18c}A. Tavakoli, Bell-type inequalities for arbitrary noncyclic networks, \textit{Phys. Rev. A} \textbf{93}, 030101(R) (2016).

\bibitem{18d}P. Contreras-Tejada, C. Palazuelos and J. I. de Vicente, Genuine multipartite nonlocality is intrinsic to quantum networks. \textit{Phys. Rev. Lett.} \textbf{126}, 040501 (2021).

\bibitem{18e}A. Tavakoli, N. Gisin and C. Branciard, Bilocal Bell inequalities violated by the quantum elegant joint measurement, arXiv:2006.16694(2021).

\bibitem{18f}M.-O. Renou and S. Beigi, Network nonlocality via rigidity of token-counting and color-matching, arXiv:2011.02769 (2021).

\bibitem{19}   J. Barrett, A. Kent and  S. Pironio, Maximally nonlocal and monogamous quantum correlations. \textit{Phys. Rev. Lett}. \textbf{97}, 170409 (2006).

\bibitem{20}  L. Masanes, S. Pironio and  A. Acin, Secure device-independent quantum key distribution with causally independent measurement devices. \textit{Nat. Commun.} \textbf{2}, 238 (2011).

\bibitem{21}  L. Aolita, R. Gallego, A.  Cabello, and A. Ac\'{\i}n, Fully nonlocal, monogamous, and random genuinely multipartite quantum correlations. \textit{Phys. Rev. Lett.} \textbf{108}, 100401 (2012).


\bibitem{22}C. M. Lee and  M. J. Hoban,  Towards device-independent information processing on general quantum networks. \textit{Phys. Rev. Lett}. \textbf{120}, 020504 (2018).

\bibitem{23}M. X. Luo,  Nonsignaling causal hierarchy of general multisource networks. \textit{Phys. Rev. A} \textbf{101}, 062317 (2020).

\bibitem{26}G. Svetlichny, Distinguishing three-body from two-body nonseparability by a Bell-type inequality. \textit{Phys. Rev. D} \textbf{35}, 3066-3069 (1987).

\bibitem{24}M.-O. Renou, E.  B\"{a}umer, S.  Boreiri, N. Brunner, N.  Gisin and  S. Beigi, Genuine quantum nonlocality in the triangle network. \textit{Phys. Rev. Lett.} \textbf{123}, 140401 (2019).


\bibitem{25}J. Pearl,  \textit{Causality}, Cambridge University Press, NY, 2009.


\bibitem{SI}Supplementary proofs of the inequalities (\ref{eq5})-(\ref{eq15}) including refs.\cite{24,26,Wrner,Graph,30,32,34,36,51,31,52,28,33,39,29,35,MBNC}.

\bibitem{Wrner} R. F. Werner, Quantum states with Einstein-Podolsky-Rosen correlations admitting a hidden-variable model. \textit{Phys. Rev. A} \textbf{40}, 4277(1989).

\bibitem{34}B. F. Toner  and   D. Bacon, Communication cost of simulating Bell correlations. \textit{Phys. Rev. Lett}. \textbf{91}, 187904 (2003).

\bibitem{36}C. Branciard, N. Brunner, H. Buhrman, R. Cleve, N. Gisin, S. Portmann, D. Rosset and M. Szegedy, Classical simulation of entanglement swapping with bounded communication. \textit{Phys. Rev. Lett}. \textbf{109}, 100401 (2012).

\bibitem{51}M. Navascues, S. Pironio  and  A. Ac\'{\i}n, Bounding the set of quatum correlations. \textit{Phys. Rev. Lett.} \textbf{98}, 010401 (2007)

\bibitem{31} J. F. Clauser, M. A. Horne, A. Shimony and  R. A. Holt, Proposed experiment to test local hidden-variable theories. \textit{Phys. Rev. Lett.} \textbf{23}, 880-884 (1969).

\bibitem{52} R. Rabelo, M. Ho, D. Cavalcanti,  N. Brunner and  V. Scarani, Device-independent certification of entangled measurements.  \textit{Phys. Rev. Lett.} \textbf{107}, 050502 (2011).


\bibitem{28} M. Zwerger,  W.  D\"{u}r,  J. D.  Bancal and  P. Sekatski, Device-independent detection of genuine multipartite entanglement for all pure states. \textit{Phys. Rev. Lett.} \textbf{122}, 060502 (2019).

\bibitem{30}  D. M. Greenberger,  M. A.  Horne and   A. Zeilinger, Going beyond Bell's Theorem, in \textit{Bell's Theorem, Quantum Theory, and Conceptions of the Universe}, M. Kafatos, ed. (Kluwer Academic, Dordrecht, 1989), p. 69.

\bibitem{32}N. D. Mermin,  Extreme quantum entanglement in a superposition of macroscopically distinct states. \textit{Phys. Rev. Lett.} \textbf{65}, 1838 (1990).

\bibitem{29} M. Zukowski,  C. Brukner,  W. Laskowski \&  M. Wiesniak, Do all pure entangled states violate Bell's inequalities for correlation functions? \textit{Phys. Rev. Lett.} \textbf{88}, 210402 (2002).


\bibitem{33}  J. Barrett,  R.  Colbeck and   A. Kent, Unconditionally secure device-independent quantum key distribution with only two devices. \textit{Phys. Rev. A} \textbf{86}, 062326 (2012).


\bibitem{35}  S. L. Braunstein and C. M. Caves,  Wringing out beter Bell inequalities.  \textit{Ann. Phys}. \textbf{202}, 22-56 (1990).

\bibitem{8s}M. Zukowski,   A. Zeilinger,   M. A.  Horne, and A. K.  Ekert, ``Event-ready-detectors'' Bell experiment via entanglement swapping.  \textit{Phys. Rev. Lett.} \textbf{71}, 4287 (1993).

\bibitem{9s}N. Gisin  and  A. Peres, Maximal violation of Bell's inequality for arbitrarily large spin. \textit{Phys. Lett. A} \textbf{162}, 15-17(1992).


\bibitem{6s} A. Ac\'{i}n,  S. Massar, and  S.  Pironio, Randomness versus nonlocality and entanglement. \textit{Phys. Rev. Lett.} \textbf{108}, 100402 (2012).


\bibitem{10s}  A.  Coladangelo,  K. T. Goh, and  V.  Scarani, All pure bipartite entangled states can be self-tested. \textit{Nat. Commun}. \textbf{8}, 15485 (2017).


\bibitem{12s}  M. Gachechiladze,   C. Budroni, and  O.  G\"{u}hne, Extreme violation of local realism in quantum hypergraph states. \textit{Phys. Rev. Lett.} \textbf{116}, 070401 (2016).
    
\bibitem{4s}  G. Svetlichny, Distinguishing three-body from two-body nonseparability by a Bell-type inequality. \textit{Phys. Rev. D} \textbf{35}, 3066-3069 (1987).

\bibitem{39}R. Raussendorf,  D. E. Browne and  H. J.  Briegel, Measurement-based quantum computation on cluster states. \textit{Phys. Rev. A} \textbf{68}, 022312 (2003).

\bibitem{MBNC}M. G. M. Moreno, S. Brito, R. V. Nery and R. Chaves, Device-independent secret sharing and a stronger form of Bell nonlocality, \textit{Phys. Rev. A} \textbf{101}, 052339 (2020).

\bibitem{38} R. Koenig, R. Renner  and  C.  Schaffner, The operational meaning of min and max-entropy. \textit{IEEE Trans. Inf. Theory} \textbf{24}, 339-348(2009).

 \bibitem{40}T.-C. Wei,  I.  Affleck and   R. Raussendorf, Affleck-Kennedy-Lieb-Tasaki State on a Honeycomb lattice is a universal quantum computational resource. \textit{Phys. Rev. Lett.} \textbf{106}, 070501 (2011).

\bibitem{41}I. Affleck,  T. Kennedy,   E. H. Lieb and   H. Tasaki, Valence bond ground states in isotropic quantum antiferromagnets. \textit{Phys. Rev. Lett}. \textbf{59}, 799 (1987).

\bibitem{Graph}R. Diestel, \textit{Graph Theory}, Springer-Verlag Berlin Heidelberg, 2017.

\bibitem{42}  H. P. Nautrup and  T. C. Wei, Symmetry-protected topologically ordered states for universal quantum computation. \textit{Phys. Rev. A } \textbf{92}, 052309 (2015).

 \bibitem{43}A. Childs,  Secure assisted quantum computation. \textit{Quantum Inform. Comput.} \textbf{5}, 456-466 (2001).

\bibitem{44} A.  Broadbent,  J.  Fitzsimons and   E. Kashefi, Universal blind quantum computation,  \textit{Proc. 50th IEEE Sympo. Found. of Comput. Sci. (FOCS)} 517-526 (2009).

\bibitem{45} B. W. Reichardt, F. Unger, and  U. Vazirani,  Classical command of quantum systems. \textit{Nature} \textbf{496}, 456-460 (2013).

\bibitem{47}  M. Hajdu\v{s}ek,  C. A.  Perez-Delgado and J. F. Fitzsimons, Device-independent verifiable blind quantum computation.  arXiv:1502.02563v2 (2015).

\bibitem{48} T.  Morimae and  K. Fujii,  Secure entanglement distillation for double-server blind quantum computation. \textit{Phys. Rev. Lett.} \textbf{111}, 020502 (2013).

\bibitem{49} A. Shamir, How to share a secret. \textit{Commun. ACM} \textbf{22}, 612-613 (1979).

\bibitem{50} M. Hillery,  V. Buzek and  A. Berthiaume, Quantum secret sharing. \textit{Phys. Rev. A} \textbf{59}, 1829-1834 (1999).

\bibitem{54}I. Ekeland  and  R. Temam, \textit{Convex analysis and variational problems}. Vol.28, SIAM, 1999.

\bibitem{55} J. W.  Hall, Local deterministic model of singlet state correlations based on relaxing measurement independence. \textit{Phys. Rev. Lett.} \textbf{105}, 250404 (2010).

\bibitem{56}G. P\"{u}tz,  D. Rosset,  T. J. Barnea,  Y. C.  Liang and N. Gisin, Arbitrarily small amount of measurement independence is sufficient to manifest quantum nonlocality. \textit{Phys. Rev. Lett.} \textbf{113}, 190402 (2014).


\bibitem{57} J. Gallicchio,  A. S.  Friedman and D. I. Kaiser,  Testing Bell's inequality with Cosmic photons: Closing the setting-independence loophole. \textit{Phys. Rev. Lett.} \textbf{112}, 110405 (2014).

\bibitem{58} J. Handsteiner,  {\it et al.} Cosmic Bell test: measurement settings from Milky way stars. \textit{Phys. Rev. Lett.} \textbf{118}, 60401 (2017).
    

\end{thebibliography}
\end{document}